\documentclass[12pt]{article}
\usepackage{epsfig,graphicx}
\usepackage[usenames]{color}

\begin{document}

\begin{center}

\vspace*{1.0cm}

{\Large \bf{Final results of the Aurora experiment to study 2$\beta$ decay of $^{116}$Cd with enriched $^{116}$CdWO$_4$ crystal scintillators}}

\vskip 1.0cm

{\bf A.S.~Barabash$^{a}$,
 P.~Belli$^{b,c}$,
 R.~Bernabei$^{b,c,}$\footnote{Corresponding author. {\it E-mail
address:} rita.bernabei@roma2.infn.it (R.~Bernabei).},
 F.~Cappella$^{d}$,
 V.~Caracciolo$^{e}$,
 R.~Cerulli$^{b,c}$,
 D.M.~Chernyak$^{f,g}$,
 F.A.~Danevich$^{f}$,
 S.~d'Angelo$^{b,c,\dagger}$,
 A.~Incicchitti$^{d,h}$,
 D.V.~Kasperovych$^{f}$,
 V.V.~Kobychev$^{f}$,
 S.I.~Konovalov$^{a}$,
 M.~Laubenstein$^{e}$,
 D.V.~Poda$^{f,i}$,
 O.G.~Polischuk$^{f}$,
 V.N.~Shlegel$^{j}$,
 V.I.~Tretyak$^{f}$,
 V.I.~Umatov$^{a}$,
 Ya.V.~Vasiliev$^{j}$ }

\vskip 0.3cm

 $^{a}${\it National Research Centre ``Kurchatov  Institute'', Institute of Theoretical
    and Experimental Physics, 117218 Moscow, Russia}

 $^{b}${\it INFN, sezione di Roma ``Tor Vergata'', I-00133 Rome, Italy}

 $^{c}${\it Dipartimento di Fisica, Universit\`{a} di Roma ``Tor Vergata'', I-00133 Rome, Italy}

 $^{d}${\it INFN, sezione di Roma, I-00185 Rome, Italy}

 $^{e}${\it INFN, Laboratori Nazionali del Gran Sasso, I-67100 Assergi (AQ), Italy}

 $^{f}${\it Institute for Nuclear Research, 03028 Kyiv, Ukraine}

 $^{g}${\it Kavli Institute for the Physics and Mathematics of the Universe, University of Tokyo, Kashiwa, 277-8583, Japan}

 $^{h}${\it Dipartimento di Fisica, Universit\`{a} di Roma ``La Sapienza'', I-00185 Rome, Italy}

 $^{i}${\it CSNSM, Univ. Paris-Sud, CNRS/IN2P3, Universit\'e Paris-Saclay, 91405 Orsay, France}

 $^{j}${\it {Nikolaev Institute of Inorganic Chemistry, 630090 Novosibirsk, Russia}

 $\dagger${Deceased} }
\end{center}

\vskip 0.5cm

\begin{abstract}

The double-beta decay of $^{116}$Cd has been investigated with the
help of radiopure enriched $^{116}$CdWO$_4$ crystal scintillators
(mass of 1.162 kg) at the Gran Sasso underground laboratory. The
half-life of $^{116}$Cd relatively to the $2\nu2\beta$ decay to
the ground state of $^{116}$Sn was measured with the highest
up-to-date accuracy as
$T_{1/2}=(2.63^{+0.11}_{-0.12})\times10^{19}$ yr. A new improved
limit on the 0$\nu$2$\beta$ decay of $^{116}$Cd to the ground
state of $^{116}$Sn was set as $T_{1/2}\geq 2.2 \times 10^{23}$ yr
at 90\% C.L., which is the most stringent known restriction for
this isotope. It corresponds to the effective Majorana neutrino
mass limit in the range $\langle m_\nu\rangle\le(1.0-1.7)$ eV,
depending on the nuclear matrix elements used in the estimations.
New improved half-life limits for the 0$\nu$2$\beta$ decay with
majoron(s) emission, Lorentz-violating $2\nu2\beta$ decay and
$2\beta$ transitions to excited states of $^{116}$Sn were set at
the level of $T_{1/2}\geq 10^{20}-10^{22}$ yr. New limits for the
hypothetical lepton-number violating parameters (right-handed
currents admixtures in weak interaction, the effective
majoron-neutrino coupling constants, R-parity violating parameter,
Lorentz-violating parameter, heavy neutrino mass) were set.

\end{abstract}

\vskip 0.2cm


\vskip 0.2cm


\section{INTRODUCTION}

The double-beta (2$\beta$) decay is a transformation of nucleus
$(A,Z)$ into $(A,Z+2)$ with simultaneous emission of two
electrons. Two-neutrino double-beta (2$\nu$2$\beta$) decay, the
process allowed in the Standard Model of particle physics (SM), is
the rarest nuclear decay ever observed (with the half-lives in the
range $T_{1/2} \simeq 10^{18} - 10^{24}$ yr
\cite{Tretyak:2002,Saakyan:2013,Barabash0:2015}). Neutrinoless
double-beta ($0\nu2\beta$) decay is forbidden in the SM because it
violates the lepton number by two units and is possible if
neutrino is a massive Majorana particle. Therefore, the investigation
of the decay is capable to clarify many questions of neutrino and
weak interaction physics: to check the lepton number conservation,
to determine the neutrino nature (Dirac or Majorana particle),
to estimate an absolute scale of the neutrino mass and the neutrino
mass hierarchy, to probe the existence of the right-handed currents
in the weak interaction, existence of majorons, to test many
extensions of the SM
\cite{Vergados:2016,Pas:2015,Bilenky:2016,Delloro:2016}. After the
seventy years of searches, the $0\nu2\beta$ decay is still not
observed, the most sensitive experiments give only limits on the
0$\nu$2$\beta$ decay half-lives for several nuclei at the level of
$\lim T_{1/2}\sim 10^{24}-10^{26}$ yr. Limits on the effective
Majorana neutrino mass of the electron neutrino on the level of
$\lim \langle m_{\nu}\rangle \sim 0.1-0.7$ eV can be obtained by
using theoretical calculations of the decay probability (see
reviews
\cite{Vergados:2016,Bilenky:2016,Delloro:2016,Giuliani:2012,Cremonesi:2014,Sarazin:2015}
and recent results
\cite{Arnold:2015,Gando:2016,Albert:2018,Alduino:2018,Aalseth:2018,Agostini:2018,Azzolini:2018}).

Experimental investigations of the $2\nu2\beta$ decay may test the
theoretical calculations of the nuclear matrix elements (NMEs) for
the $0\nu2\beta$ decay processes \cite{Engel:2017}. In particular,
precise measurements of 2$\nu$2$\beta$ decay rate for different
nuclei can help to solve problem of the axial vector coupling
constant $g_A$ value (see discussions in
\cite{Vergados:2016,Barea:2015,Kostensalo:2017}), while accurate
investigation of the $2\nu2\beta$ decay spectral shape can help to
determine the mechanism of decay (high state dominance or single
state dominance \cite{Abad:1984}), to test existence of
hypothetical bosonic neutrinos \cite{Barabash:2007}, and to check
Lorentz and CPT violation \cite{Diaz:2013}.

\begin{figure}[!ht]
 \begin{center}
 \resizebox{0.5\textwidth}{!}{\includegraphics{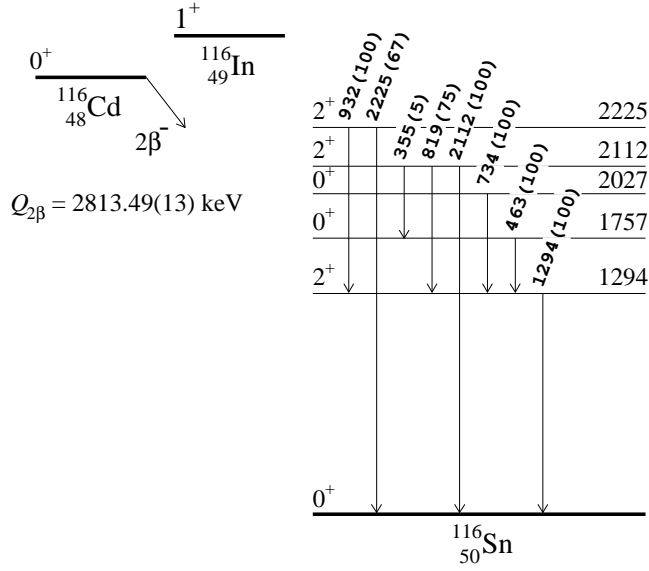}}
 \end{center}
 \caption{Simplified decay scheme of $^{116}$Cd
\cite{Blachot:2010}. Energies of the excited levels and emitted
$\gamma$ quanta are in keV. The relative intensities of $\gamma$
quanta are given in parentheses.}
 \label{fig:01}
 \end{figure}

The nuclide $^{116}$Cd is one of the most favorable candidates for
the $0\nu2\beta$ searches thanks to the high energy of decay
($Q_{2\beta}=2813.49(13)$ keV \cite{Wang:2017}), the promising
estimations of the decay probability
\cite{Barea:2015,Simkovic:2013,Vaquero:2013,Hyvarinen:2015,Yao:2015,Song:2017},
a relatively large isotopic abundance ($\delta=7.512(54)\%$
\cite{Meija:2016}), the availability of enrichment by
ultra-centrifugation in large amount, possibilities to realize a
calorimetric ``source = detector'' experiment with cadmium
tungstate (CdWO$_4$) crystal scintillators already successfully
used in several low counting experiments
\cite{Danevich:2003a,Danevich:2003b,Belli:2007,Belli:2008,Belli:2012,Belli:2016}.
$^{116}$Cd is considered as one of the most promising nuclei for a
large scale bolometric experiment CUPID to explore the inverted
hierarchy of the neutrino mass pattern \cite{CUPID,Giuliani:2018}.
A simplified scheme of $^{116}$Cd $2\beta$ decay is shown in Fig.
\ref{fig:01}.

\begin{table*}[!ht]
 \caption{Experiments where $2\nu2\beta$ decay of $^{116}$Cd was observed.}
\begin{center}
\begin{tabular}{llll}
\hline
 Experiment                                         & $T_{1/2} (\times 10^{19}$ yr)                                     & Year, Reference \\
\hline
 ELEGANT V, $^{116}$Cd foil, drift chambers,        &                                                                   &  \\
 plastic scintillators                              & $2.6^{+0.9}_{-0.5}$                                               &  1995 \cite{Ejiri:1995} \\
\hline
 Solotvina, $^{116}$CdWO$_4$ scintillators          & $2.7^{+0.5}_{-0.4}(\mathrm{stat})^{+0.9}_{-0.6} (\mathrm{sys})$  & 1995 \cite{Danevich:1995} \\
\hline
 NEMO-2, $^{116}$Cd foils, track reconstruction     &                                                                   &  \\
 by Geiger cells,  plastic scintillators            &  $3.75\pm0.35(\mathrm{stat})\pm 0.21(\mathrm{sys})^{a}$           & 1995 \cite{NEMO2:1995,NEMO2:1996} \\
\hline
 Solotvina, $^{116}$CdWO$_4$ scintillators          &  $2.6\pm 0.1(\mathrm{stat})^{+0.7}_{-0.4}(\mathrm{sys})$          & 2000 \cite{Danevich:2000} \\
\hline
 Solotvina, $^{116}$CdWO$_4$ scintillators          & $2.9\pm 0.06(\mathrm{stat})^{+0.4}_{-0.3}(\mathrm{sys})$          &  2003 \cite{Danevich:2003b} \\
\hline
 NEMO-3, $^{116}$Cd foils, track reconstruction     &                                                                   &  \\
 by Geiger cells, plastic scintillators             &  $2.74\pm0.04(\mathrm{stat})\pm 0.18(\mathrm{sys})$               & 2017 \cite{NEMO3:2017} \\
\hline
 $^{116}$CdWO$_4$ scintillators                     &  $2.63\pm0.01(\mathrm{stat})^{+0.11}_{-0.12}(\mathrm{sys})$       & 2018, Present work \\
\hline
 \multicolumn{4}{l}{$^{a)}$ The result of NEMO-2 was re-estimated as $T_{1/2}=[2.9\pm0.3(\mathrm{stat})\pm0.2(\mathrm{sys})]\times10^{19}$ yr in \cite{NEMO2*:2010}.} \\
\end{tabular}
\end{center}
\label{tab:2n2bres}
\end{table*}

The process of two-neutrino $2\beta$ decay of $^{116}$Cd was
observed for the first time in the ELEGANT V experiment
\cite{Ejiri:1995} at the Kamioka underground laboratory with the
half-life $2.6^{+0.9}_{-0.5}\times 10^{19}$~yr by using drift
chambers and plastic scintillators to measure electrons emitted in
the decay (see Table \ref{tab:2n2bres} where the positive results
of $2\nu2\beta$ studies are presented). Then the decay was
observed in the calorimetric experiment at the Solotvina
underground laboratory with cadmium tungstate crystal
scintillators enriched in the isotope $^{116}$Cd
\cite{Danevich:2003b,Danevich:1995,Danevich:2000}. The decay was
also detected by the NEMO-2 and NEMO-3 tracking set-ups
\cite{NEMO2:1995,NEMO2:1996,NEMO3:2017}. The last experiment gives
up-to-date the most accurate value of the half-life
$T_{1/2}=2.74\pm0.04(\mathrm{stat})\pm
0.18(\mathrm{sys})\times10^{19}$ yr \cite{NEMO3:2017}.

The most stringent limit on $0\nu2\beta$ decay of $^{116}$Cd
($T_{1/2}\geq1.7\times10^{23}$ yr at 90\% confidence level, C.L.)
was set in the Solotvina experiment \cite{Danevich:2003b}. A
similar half-life limit was obtained recently by the NEMO-3
collaboration as $T_{1/2}\geq1.0\times10^{23}$ yr at 90\% C.L.
\cite{NEMO3:2017}. The most sensitive searches for $2\beta$
transitions to excited levels of $^{116}$Sn, and for $0\nu2\beta$
decay with majorons emission have been also realized in the
Solotvina experiment with the half-life limits on the level of
$T_{1/2}\geq 10^{20}-10^{22}$ yr. The $0\nu2\beta$ decay with
majoron emission was investigated by the NEMO-3 collaboration too
\cite{NEMO3:2017}. The $2\beta$ transitions to excited levels were
also searched for by low-background $\gamma$ spectrometry with
high-purity germanium detectors \cite{Barabash:1990,Piepke:1994}.

Here we report the final results of the Aurora experiment to study
different modes and channels of 2$\beta$ decay of $^{116}$Cd
performed in $2011-2017$ at the Gran Sasso underground laboratory
with the help of more than 1 kg radiopure $^{116}$CdWO$_4$ crystal
scintillators enriched in the isotope $^{116}$Cd. Preliminary
results of the experiment were reported in the conference
proceedings
\cite{Barabash:2013,Poda:2014,Polischuk:2015,Danevich:2016,Polischuk:2017}.

\section{EXPERIMENT}

\begin{figure}[!ht]
 \begin{center}
 \resizebox{0.54\textwidth}{!}{\includegraphics{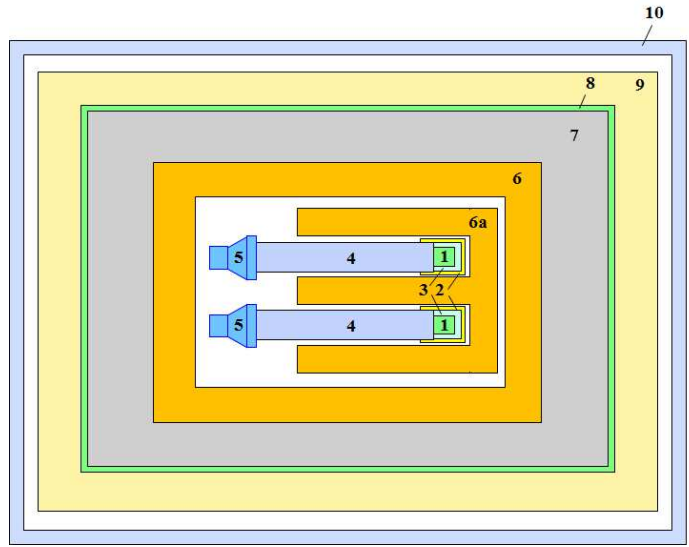}}
 \end{center}
 \caption{Schematic cross-sectional view of the Aurora set-up. There were
$^{116}$CdWO$_4$ crystal scintillators (1) fixed in Teflon
containers (2) filled up with liquid scintillator (3) and viewed
through quartz light-guides (4) by photomultipliers (5). The
passive shield consisted of high purity copper (6), additional
high purity copper shield (6a), low radioactive lead (7), cadmium
(8), polyethylene/paraffin (9), plexiglas box (10).}
 \label{fig:set-up}
 \end{figure}

Two cadmium tungstate crystals (580 g and 582 g, denoted here as
No.~1 and No.~2, respectively) produced with the help of the
low-thermal-gradient Czochralski crystal growth technique from
highly purified cadmium enriched in $^{116}$Cd to 82\%
\cite{Barabash:2011} were used for the investigations of 2$\beta$
decay of $^{116}$Cd. The experiments have been realized in the low
background DAMA/R\&D set-up installed deep underground
($\approx3600$~m~w.e.) at the Gran Sasso laboratory of I.N.F.N.
(Italy). There were several upgrades of the experimental set-up
aiming at improvement of the detector background counting rate and
energy resolution, and several studies about the crystal
scintillators radioactive contamination
\cite{Poda:2014,Barabash:2011,Poda:2013,Danevich:2013}. In the
final stage of the experiment (since 18 March 2014) the
scintillators were fixed inside polytetrafluoroethylene containers
(see a schematic cross-sectional view of the Aurora set-up in Fig.
\ref{fig:set-up}) filled up with ultra-pure pseudocumene based
liquid scintillator (LS). 
The $^{116}$CdWO$_4$ crystals and the LS
were viewed through high purity quartz light guides ($\oslash$7
$\times$ 40 cm) by 3 inches low radioactive photomultiplier tubes
(PMT, Hamamatsu R6233MOD). The detector was installed inside a
passive shield assembled from high purity copper (10 cm), low
radioactive lead (15 cm), cadmium (1.5 mm) and
polyethylene/paraffin (4 to 10 cm) to reduce the external
background. The whole set-up was contained inside a plexiglas box
and continuously flushed by high purity nitrogen gas to remove
environmental radon.

An event-by-event data acquisition system (DAQ) based on a 1 GS/s
8-bit transient digitizer (Acqiris DC270) recorded the amplitude,
the arrival time and the pulse shape of each event (over 50 $\mu$s
with a time bin of 20 ns). Multiple events were acquired in a
single buffer in the DAQ program (190 events per each buffer,
without dead time). The energy scale and the energy resolution of
the detectors were measured in the beginning, several times during
the measurements, and at the end of the experiment with $^{22}$Na,
$^{60}$Co, $^{133}$Ba, $^{137}$Cs, and $^{228}$Th $\gamma$
sources. The data of the calibration measurements were used to set
a dependence of the energy resolution on energy. The energy
resolution of the detector to $\gamma$ quanta with energy
$E_{\gamma}$ can  be described by the function FWHM$_{\gamma}$ =
$\sqrt{10.2 \times E_{\gamma}}$, where FWHM$_{\gamma}$ (Full Width
at Half Maximum) and $E_{\gamma}$ are given in keV. The energy
scale during the experiment was reasonably stable with deviation
in the range of $\pm0.9\%$.

\section{DATA ANALYSIS}

The pulse-shape discrimination (PSD) between $\gamma$($\beta$) and
$\alpha$ particles, the time-amplitude analysis of fast sub-chains
of decays from the $^{232}$Th family, the front-edge analysis of
the pulse shape, and the Monte Carlo simulation of the measured
energy spectra have been applied to estimate the radioactive
contamination of the $^{116}$CdWO$_4$ crystal scintillators, the
response of the detector to $\alpha$ particles, and to reject the
detectors background. The data on radioactive contamination of the
$^{116}$CdWO$_4$ crystal scintillators were then used to build a
model of the background that is a crucial issue to estimate the
$^{116}$Cd half-life relatively to the two-neutrino mode of
$2\beta$ decay and derive limits on the $2\beta$ processes that
have not been observed.

\subsection{Pulse-shape discrimination between $\gamma$($\beta$) and $\alpha$ particles}
\label{sec:psd}

\begin{figure}[ht]
 \begin{center}
 \resizebox{0.54\textwidth}{!}{\includegraphics{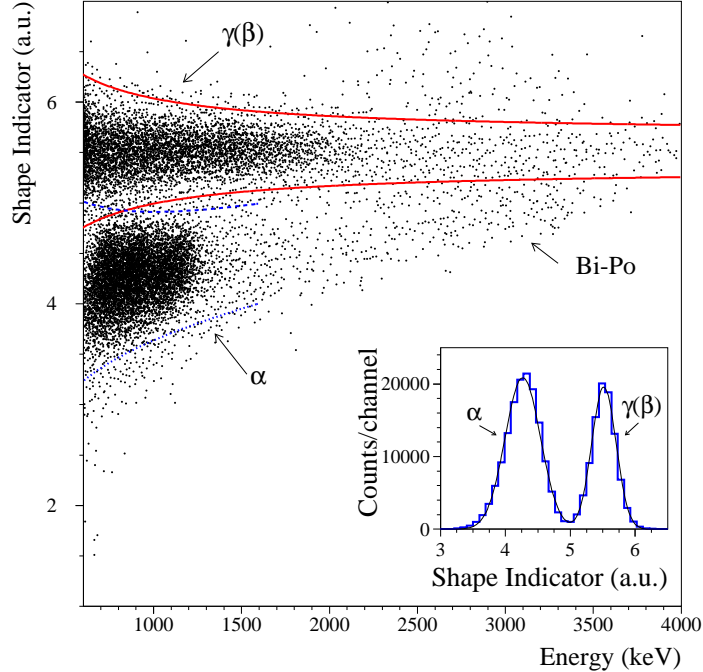}}
 \end{center}
 \caption{Shape indicator (see text) versus energy for the background
data accumulated over 26831 h with the $^{116}$CdWO$_4$ crystal
scintillator No.~2. The 2.33-sigma intervals (98\% of events) for
the shape indicator values corresponding to $\gamma$($\beta$) and
$\alpha$ particles are depicted by solid and dotted lines,
respectively. The population of events in the energy interval
$\sim(1.7-4)$ MeV with shape indicator values between $\sim(4-7)$
are caused by the decays of the fast $^{212}$Bi$-^{212}$Po
sub-chain of the $^{232}$Th chain. (Inset) Distribution of shape
indicators for the events with the energies in the range of
$0.7-1.4$ MeV. The fit of the distribution by Gaussian functions
is shown by solid line.}
 \label{fig:02}
 \end{figure}

\begin{figure}[ht]
 \begin{center}
 \resizebox{0.54\textwidth}{!}{\includegraphics{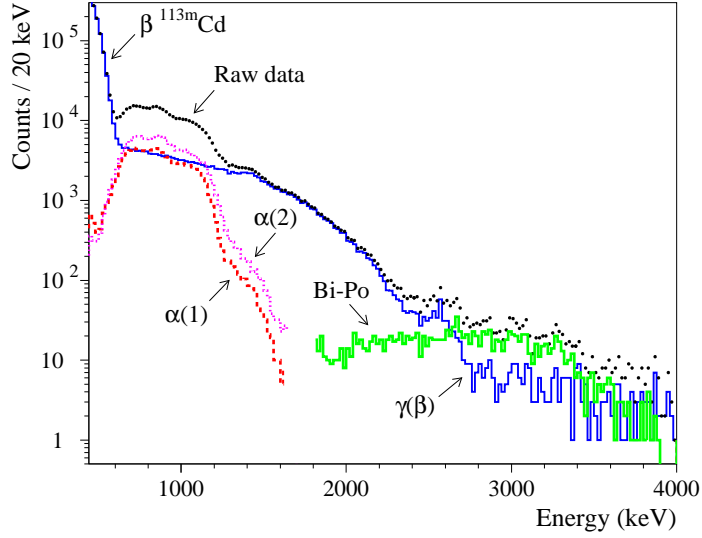}}
 \end{center}
 \caption{The sum energy spectrum acquired with two
$^{116}$CdWO$_4$ detectors over 26831 h (Raw data) and spectra of
$\gamma(\beta)$, $\alpha$ and $^{212}$Bi$~-^{212}$Po events
(denoted ``Bi-Po'') selected by the pulse-shape and the front-edge
analyzes described in text (for the front-edge analysis see
Section \ref{sec:front}). The spectra $\alpha$(1) and $\alpha$(2)
denote the distributions of alpha events accumulated by the
detectors No.~1 and No.~2, respectively.}
 \label{fig:03}
 \end{figure}

The optimal filter method proposed by E. Gatti and F. De Martini
\cite{Gatti:1962}, developed for CdWO$_4$ scintillation detectors
\cite{Fazzini:1998,Bardelli:2006}, was applied to analyze the
pulse profiles of the events aiming at discrimination of
$\gamma$($\beta$) events from those induced by $\alpha$ particles.
For each signal $f(t)$, the numerical characteristic of its shape
(shape indicator, $SI$) was defined by using the following
equation:

\begin{equation}
SI = \sum f(t_{k})  \times P(t_{k}) / \sum f(t_{k}),
\end{equation}

\noindent where the sum is over the time channels $k$, starting
from the origin of signal up to 50 $\mu$s; $f(t_{k})$ is the
digitized amplitude (at the time $t_{k}$) of a given signal. The
weight function $P(t)$ was defined as:

\begin{equation}
P(t) = |f_{\alpha}(t)-f_{\gamma}(t)|/| f_{\alpha}(t)+f_{\gamma}(t)|,
\end{equation}

\noindent where $f_{\alpha}(t)$ and $f_{\gamma}(t)$ are the
reference pulse shapes for $\alpha$ particles and $\gamma$ quanta,
respectively. By using this approach, $\alpha$ events were clearly
separated from $\gamma(\beta)$ events. The scatter plot of the
shape indicator versus energy for the data of the low background
measurements is shown in Fig. \ref{fig:02}; it demonstrates the
pulse-shape discrimination ability of the $^{116}$CdWO$_4$
detector. The distribution of shape indicators for the events with
the energies in the range of $0.7-1.4$ MeV is shown in Inset of
Fig. \ref{fig:02}. The spectra of $\gamma(\beta$) and $\alpha$
events selected by the pulse-shape analysis are presented in  Fig.
\ref{fig:03}. The total alpha activity of U/Th with their
daughters in the crystal  No.~2 is higher than that in the crystal
No.~1 due to segregation of impurities (particularly of
radioactive elements) in the crystal growth process
\cite{Barabash:2016}. The total internal $\alpha$ activity in the
crystals No.~1 and No.~2 is 1.8(2) mBq/kg and 2.7(3) mBq/kg,
respectively.

 \begin{figure}[h!]
 \begin{center}
 \resizebox{0.5\textwidth}{!}{\includegraphics{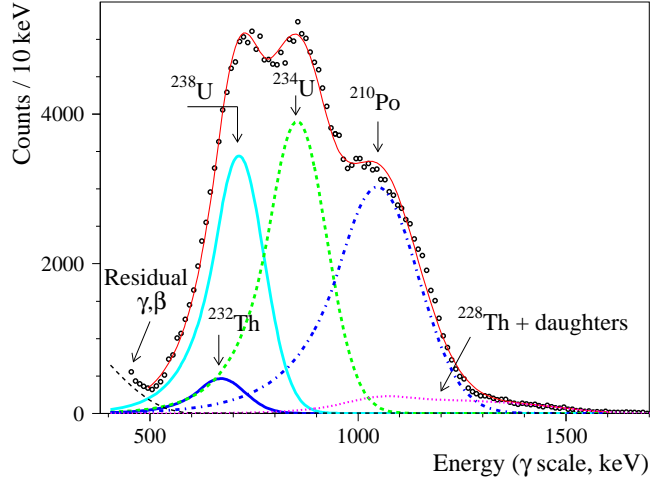}}
 \caption{The sum energy spectrum of $\alpha$ events (points) selected by the
pulse-shape discrimination from the data of low-background
measurements with the $^{116}$CdWO$_4$ crystal scintillators No.~1
and No.~2 over 26831 h. The fit of the data by the model built from
$\alpha$ decays of $^{238}$U and $^{232}$Th with daughters, and
residual $\gamma, \beta$ background is shown by solid line (the
individual components of the fit are shown too).}
 \label{fig:04}
\end{center}
\end{figure}

A sum $\alpha$ energy spectrum of the two detectors (see Fig.
\ref{fig:04}) was fitted by using a model which includes $\alpha$
peaks of $^{232}$Th, $^{238}$U and their daughters, plus
$\gamma/\beta$ background. The equilibrium of the $^{232}$Th and
$^{238}$U chains is assumed to be broken in the $^{116}$CdWO$_4$
crystals. Therefore, activities of $^{238}$U, $^{234}$U,
$^{230}$Th, $^{226}$Ra, $^{210}$Po, $^{232}$Th, $^{228}$Th were
free parameters of the fit. We have found that the spectral shape
of the individual alpha peaks is described better by using
the asymmetrical function proposed in \cite{Koskelo:1996} (see Section
\ref{sec:t-a}). The energy resolution of the detector to $\alpha$
particles and the $\alpha$/$\gamma$ ratio\footnote{The
$\alpha$/$\gamma$ ratio is the light yield of $\alpha$ particles
$LY_{\alpha}$ divided by light yield of gamma quanta $LY_\gamma$
of the same energy. Because of quenching, $LY$ for heavy particles
in scintillators is lower than that for electrons (or $\gamma$
quanta), depending on the particle's type and energy
\cite{Tretyak:2010}.} were taken as free parameters of the fit.
Furthermore, we should use different $\alpha$/$\gamma$ ratio and
energy resolution to describe peak of $^{210}$Po in the spectrum,
that can be explained by non-uniform distribution of different
impurities in the crystals volume, particularly of $^{210}$Pb that
is parent nuclide for $^{210}$Po. The effect can be explained by
two possible origins of $^{210}$Pb in the crystals: as lead
impurity (since lead always contains some amount of radioactive
$^{210}$Pb), and as product of radium decay (isotope $^{226}$Ra,
daughter of $^{238}$U). Besides, one cannot exclude surface
contamination of the crystals by $^{210}$Pb caused by decays of
radon present in air. Both $\alpha/\gamma$ ratio and energy
resolution are higher for the $^{210}$Po $\alpha$ peak than that
for other $\alpha$ active nuclides.

The result of the fit in the energy interval $(470-1600)$ keV is
shown in Fig. \ref{fig:04}. The fit gives the activities of
$^{238}$U, $^{234}$U, $^{210}$Po, $^{232}$Th and $^{228}$Th in the
crystals presented in Table \ref{tab:rc}, while only limits were
obtained for the activity of $^{230}$Th and $^{226}$Ra (the
activity of $^{228}$Th was then estimated with a higher accuracy
with the help of the time-amplitude and front-edge analyzes,
Sections \ref{sec:t-a} and \ref{sec:front}). The reference date
(February 2016) is given to take into account the decay of
$^{228}$Th (the half-life is $T_{1/2}=1.9116$ yr) and $^{110m}$Ag
($T_{1/2}= 249.83$ d)  in the crystals.

\begin{table*}[htbp]
\caption{Radioactive contamination of the $^{116}$CdWO$_4$
crystals. Reference date is February 2016.}
 \begin{center}
 \begin{tabular}{lll}
 \hline
 Chain      & Nuclide               & Activity (mBq/kg) \\
 \hline
 ~          & $^{40}$K              & $0.22(9)$ \\
  ~         & $^{90}$Sr$-^{90}$Y    & $\leq0.02$ \\
 ~          & $^{110m}$Ag           & $\leq0.007$  \\
 ~          & $^{116}$Cd            & $1.138(5)$ \\
 $^{232}$Th & $^{232}$Th            & 0.07(2) \\
 ~          & $^{228}$Ra            & $\leq0.005$ \\
 ~          & $^{228}$Th            & 0.020(1) \\
 $^{235}$U  & $^{227}$Ac            & $\leq0.002$ \\
 $^{238}$U  & $^{238}$U             & 0.58(4) \\
 ~          & $^{234}$U             & 0.6(1)\\
 ~          & $^{230}$Th            & $\leq0.13$ \\
 ~          & $^{226}$Ra            & $\leq0.006$ \\
 ~          & $^{210}$Pb            & 0.70(4) \\
 Total $\alpha$ & ~                 & 2.14(2) \\
 \hline
 \end{tabular}
 \end{center}
\label{tab:rc}
\end{table*}

\subsection{Time-amplitude analysis of fast sub-chains}
\label{sec:t-a}

\subsubsection{Selection of the $^{224}$Ra $\to$ $^{220}$Rn $\to$ $^{216}$Po $\to$ $^{212}$Pb sub-chain}
\label{sec:t-a-1}

The time-amplitude analysis (described e.g. in
\cite{Danevich:2003a,Barton:2000a,Danevich:2001}) was used to
select events of the following decay sub-chain of the $^{232}$Th
family:

\begin{center}
$^{224}$Ra ($Q_{\alpha}=5789$ keV; $T_{1/2} = 3.632$ d)
$\rightarrow$ $^{220}$Rn ($Q_{\alpha} = 6405$ keV; $T_{1/2} =
55.6$ s) $\rightarrow$ $^{216}$Po ($Q_{\alpha} = 6906$ keV;
$T_{1/2} = 0.145$ s) $\rightarrow$ $^{212}$Pb.
\end{center}

\begin{figure}[h!]
 \begin{center}
 \resizebox{0.54\textwidth}{!}{\includegraphics{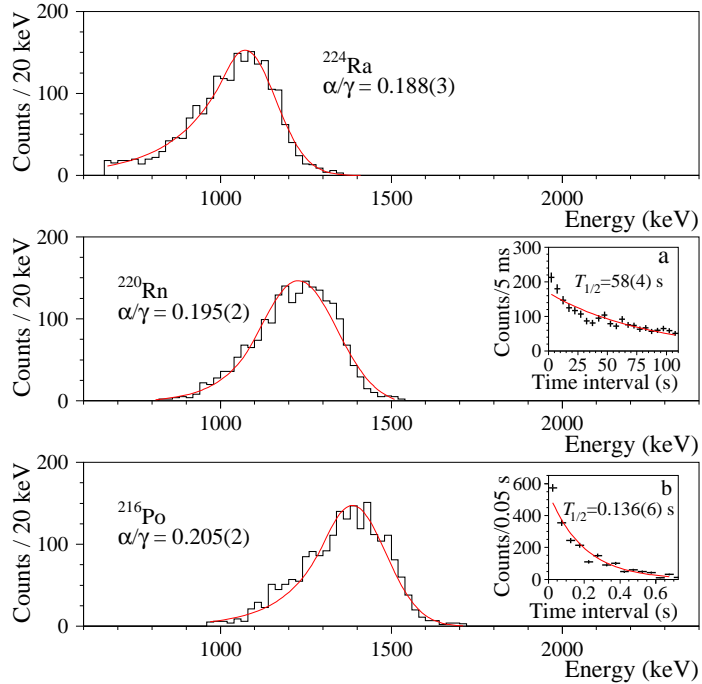}}
 \end{center}
\caption{Alpha peaks of $^{224}$Ra, $^{220}$Rn and $^{216}$Po
selected by the time-amplitude analysis from the data accumulated
over 26831 h with the $^{116}$CdWO$_4$ detectors No.~1 and No.~2.
The obtained half-lives of $^{220}$Rn ($58\pm4$ s, Inset a) and
$^{216}$Po ($0.136\pm0.006$ s, Inset b) are in agreement with the
table values ($55.6\pm0.1$ s and $0.145\pm0.002$ s,
respectively).}
 \label{fig:05}
\end{figure}

To select decays of the sub-chain, all $\alpha$ events within an
energy interval $0.82-1.54$ MeV were used as triggers ($\alpha$
particles of $^{220}$Rn), while a time interval $0-0.725$ s and
the $0.96-1.72$ MeV energy window were set for the second $\alpha$
events ($^{216}$Po). Taking into account the efficiency of the
events selection in this time interval (96.88\% of $^{216}$Po
decays), the activity of $^{228}$Th in the $^{116}$CdWO$_4$
crystals No.~1 and No.~2 was calculated as 0.013(3) mBq/kg and
0.029(4) mBq/kg, respectively. All the selected pairs
$^{220}$Rn$-$$^{216}$Po were used as triggers to find events of
$^{224}$Ra $\alpha$ decay. A $0-111$ s time interval was chosen to
select events in the energy interval $0.66-1.36$ MeV. The obtained
$\alpha$ peaks from the
$^{224}$Ra$\to$$^{220}$Rn$\to$$^{216}$Po$\to$$^{212}$Pb sub-chain
and the time distributions for the $^{220}$Rn$\to$$^{216}$Po and
$^{216}$Po$\to$$^{212}$Pb decays in the $^{116}$CdWO$_4$ detectors
No.~1 and No.~2 are shown in Fig. \ref{fig:05}. The estimated
half-lives of $^{220}$Rn and $^{216}$Po are in agreement with
those table values. An averaged activity of $^{228}$Th in the
$^{116}$CdWO$_4$ crystal scintillators estimated by using the
time-amplitude analysis is given in Table \ref{tab:rc}. It should
be stressed that the fit of the alpha spectra were performed using
non-Gaussian function for individual $\alpha$ peaks proposed in
\cite{Koskelo:1996}. The non-Gaussian shape of the  peaks can be
explained by the non-uniformity of the U/Th impurities
concentration in the crystals \cite{Barabash:2016} and, as a
result, by non-uniformity of the light collection in the
detector's volume.

No events were found with the time-amplitude analysis aiming at
search for the following fast sub-chain of the $^{235}$U family
(expected to be in equilibrium with $^{227}$Ac):

\begin{center}
$^{219}$Rn ($Q_{\alpha}=6946$ keV; $T_{1/2} = 3.96$ s)
$\rightarrow$ $^{215}$Po ($Q_{\alpha} = 7526$ keV; $T_{1/2} =
1.781$ ms) $\rightarrow$ $^{211}$Pb.
\end{center}

\noindent As a result we set a limit on activity of $^{227}$Ac in
the crystals on the level of $\leq0.002$ mBq/kg.

\subsubsection{Selection of $^{212}$Bi$\to ^{208}$Tl events}
\label{sec:t-a-2}

\begin{figure}[h!]
 \begin{center}
 \resizebox{0.5\textwidth}{!}{\includegraphics{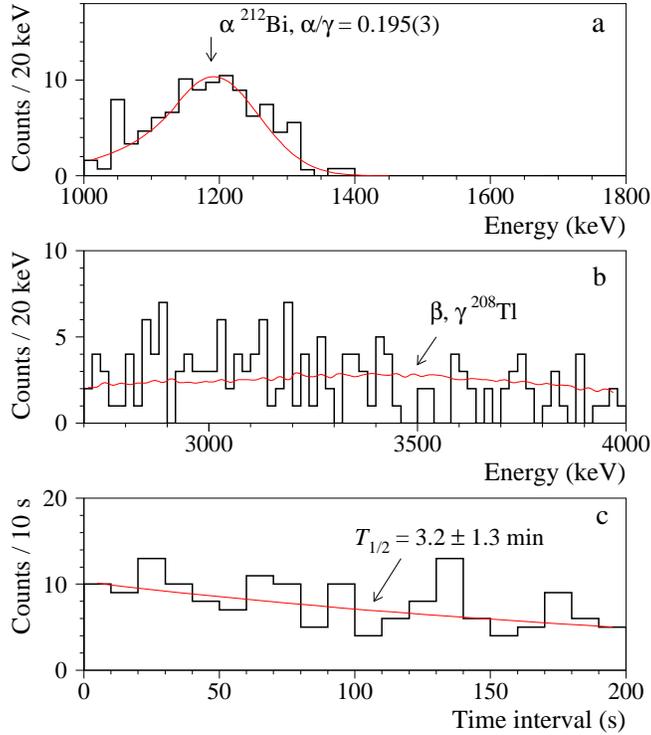}}
 \end{center}
\caption{The energy (a) and (b), and time (c) distributions for
the sequence of $\alpha$ ($^{212}$Bi) and $\beta, \gamma$
($^{208}$Tl) decays selected from the data accumulated over 26831
h with the $^{116}$CdWO$_4$ detectors No.~1 and No.~2. The fit of
the $\alpha$ peak (a), the approximation of the energy
distribution by a Monte Carlo simulated distribution of $^{208}$Tl
events (b), and the fit of the time distribution by an exponential
function with the half-life 3.2(13) minutes (c) are shown.}
 \label{fig:bi-tl}
\end{figure}

The following chain of decays: $^{212}$Bi ($Q_\alpha=6207$ keV)
$\to$ $^{208}$Tl ($Q_\beta$ = 4999 keV, $T_{1/2}$ = 3.053
min)$~\to$ $^{208}$Pb was selected by using the time-amplitude
analysis. All $\alpha$ events within the energy interval $1.0-1.4$
MeV (which contains $\alpha$ peak of $^{212}$Bi) were used as
triggers, and all the subsequent $\gamma/\beta$ events in the
energy interval $2.7-4.0$ MeV were selected within a time interval
$(0.0001-200)$ s (containing 53\% of $^{208}$Tl decays). The
capability of the analysis is demonstrated in Fig.
\ref{fig:bi-tl}. The alpha peak of $^{212}$Bi was fitted by the
asymmetric function \cite{Koskelo:1996} giving the $\alpha/\gamma$
ratio 0.195(3). The distribution of the second events is well
described by the simulated spectrum of $\beta$ and $\gamma$ events
of $^{208}$Tl, while the distribution of time intervals between
the events can be approximated by exponential function with the
half-life $3.2\pm1.3$ min, in a reasonable agreement with the
table value for $^{208}$Tl.

The selection procedure reduces the background in the high energy
part of the spectrum of $\beta$ and $\gamma$ events, however, the
procedure decreases also the live time of measurements. For this
reason, the data obtained after subtraction of the $^{208}$Tl
events were not used for estimations of double-beta processes in
$^{116}$Cd.

\begin{figure}[h!]
 \begin{center}
 \resizebox{0.54\textwidth}{!}{\includegraphics{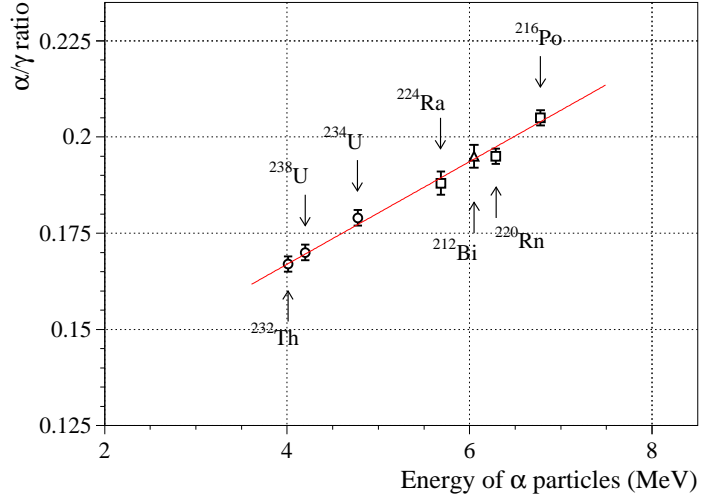}}
 \end{center}
 \vspace{-0.8cm}
\caption{Dependence of $\alpha/\gamma$ ratio on energy of the $\alpha$
particles. The open circles are from the PSD (Fig. \ref{fig:04}),
the open squares are from the time-amplitude analysis (Fig. \ref{fig:05}), and
the triangle is obtained by the analysis of the sequence $^{212}$Bi-$^{208}$Tl (Fig. \ref{fig:bi-tl}).}
 \label{fig:agr}
\end{figure}

By using positions of the $\alpha$ peaks of $^{224}$Ra, $^{220}$Rn
and $^{216}$Po (from the time-amplitude analysis, see Fig.
\ref{fig:05}), of $^{232}$Th, $^{238}$U and $^{234}$U (from the
pulse-shape discrimination, see Fig. \ref{fig:04}) and of
$^{212}$Bi (obtained by the analysis of the sequence
$^{212}$Bi$-$$^{208}$Tl presented in Fig. \ref{fig:bi-tl}), the
following dependence of $\alpha$/$\gamma$ ratio on energy of
$\alpha$ particles was obtained:
$\alpha/\gamma=0.114(7)+0.0133(12)E_{\alpha}$ in the energy
interval $4.0-6.8$ MeV ($E_{\alpha}$ is in MeV). The dependence of
the $\alpha/\gamma$ ratio on energy of $\alpha$ particles is
presented in Fig. \ref{fig:agr}.

\subsection{Discrimination of $^{212}$Bi$-$$^{212}$Po events by front-edge analysis}
\label{sec:front}

The front-edge analysis was developed to reject the following fast
sub-chain of decays from the $^{232}$Th family (Bi-Po events):

\begin{center}
$^{212}$Bi ($Q_{\beta}=2252$ keV; $T_{1/2} = 60.55$ m)
$\rightarrow$ $^{212}$Po ($Q_{\alpha} = 8954$ keV; $T_{1/2} =
0.299$ $\mu$s) $\rightarrow$ $^{208}$Pb.
\end{center}

\begin{figure}[h!]
 \begin{center}
 \resizebox{0.50\textwidth}{!}{\includegraphics{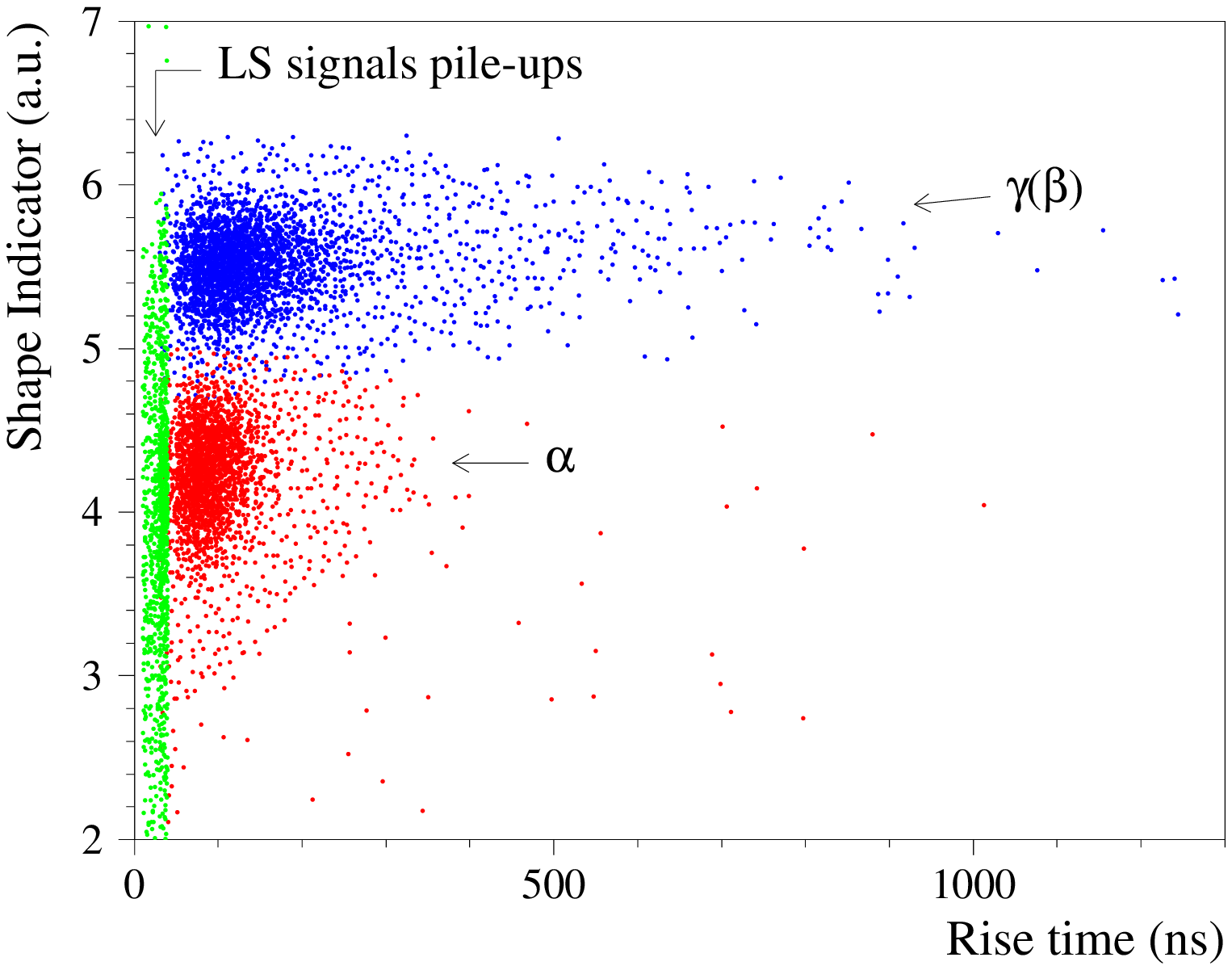}}~\resizebox{0.50\textwidth}{!}{\includegraphics{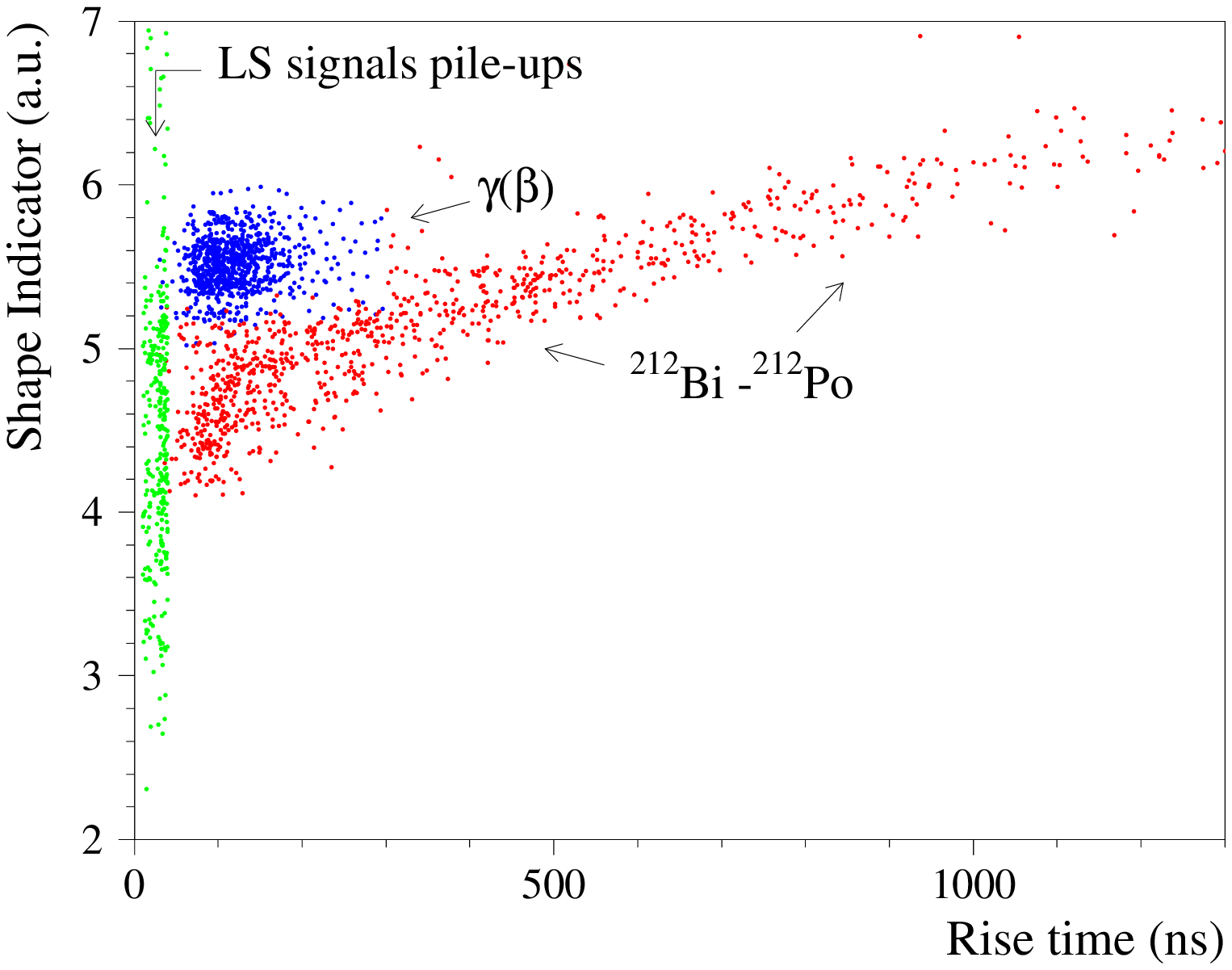}}
 \end{center}
 \vspace{-0.8cm}
 \caption{Distributions of shape indicator versus rise time for the
background events acquired with the $^{116}$CdWO$_4$ detector
No.~2 over 26831 h  in the energy interval $0.6-1.3$ MeV (left
panel) and in the energy interval $1.7-4.0$ MeV (right panel). }
\label{fig:fe}
 \end{figure}

A front-edge parameter (rise time) for each signal was calculated
as time between the signal origin and the time where signal reach
0.7 of its maximal value. Results of the front-edge analysis are
illustrated in Fig. \ref{fig:fe}, where the scatter plots of the
shape indicator versus pulse rise time for the background data
gathered over 26831 h with the $^{116}$CdWO$_4$ detector No.~2 are
shown for the events selected in the energy intervals $0.6-1.3$
MeV and $1.7-4.0$ MeV. The $1.7-4.0$ MeV data contain events with
longer rise time that is in agreement with an expected sum energy
release in the $^{212}$Bi$-^{212}$Po decay $\sim 1.8-4.4$ MeV. An
energy spectrum of the $^{212}$Bi$-^{212}$Po events selected by
the front-edge analysis is shown in Fig. \ref{fig:03}. It should
be stressed that $^{212}$Bi$-^{212}$Po events are also visible in
Fig. \ref{fig:02} since the PSD analysis is sensitive to these
events too.

The analysis allowed to estimate the activity of $^{212}$Bi (which
is in equilibrium with $^{228}$Th) in the crystals No.~1 and No.~2
as 0.018(2) mBq/kg and 0.027(3) mBq/kg, respectively, in a
reasonable agreement with the results of the time-amplitude
analysis (Section \ref{sec:t-a}). All the selected Bi-Po events
were discarded from the data that reduced background counting rate
in the energy region of interest ($2.7-2.9$ MeV) by a factor of
$\sim1.5$.

It should be stressed that the front-edge analysis rejects also
pile-ups of liquid scintillator pulses with $^{116}$CdWO$_4$
signals thanks to a shorter rise time (less than 38 ns, see Fig.
\ref{fig:fe}) of the liquid scintillator pulses.

\section{RESULTS AND DISCUSSION}

\subsection{$2\nu2\beta$ decay of $^{116}$Cd to the ground state of $^{116}$Sn}

\begin{figure}[h!]
 \begin{center}
 \resizebox{0.5\textwidth}{!}{\includegraphics{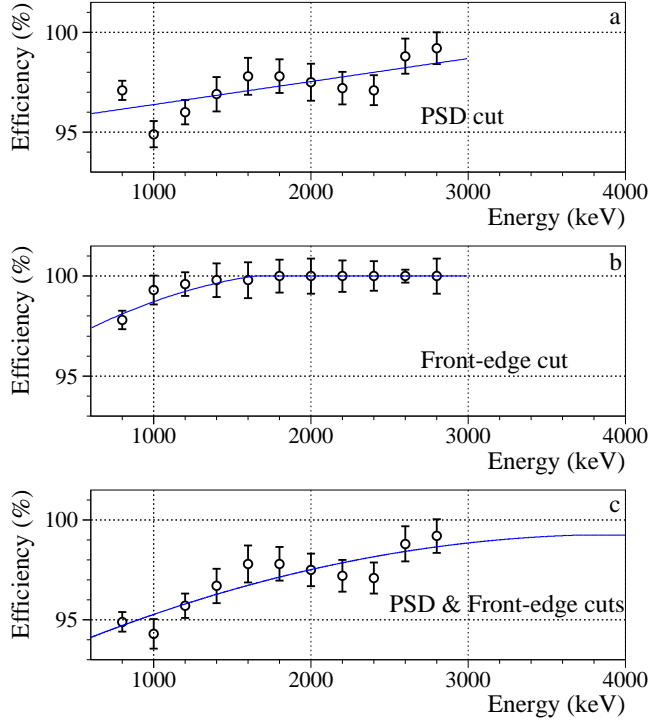}}
 \end{center}
 \vspace{-0.8cm}
\caption{
Efficiencies measured for $\gamma$ rays from $^{228}$Th calibration source when applying the pulse-shape discrimination cut (a),
the front-edge analysis cut (b) and both of them (c).}
 \label{fig:psd-fe-eff}
\end{figure}

\begin{figure}[h!]
 \begin{center}
 \resizebox{0.54\textwidth}{!}{\includegraphics{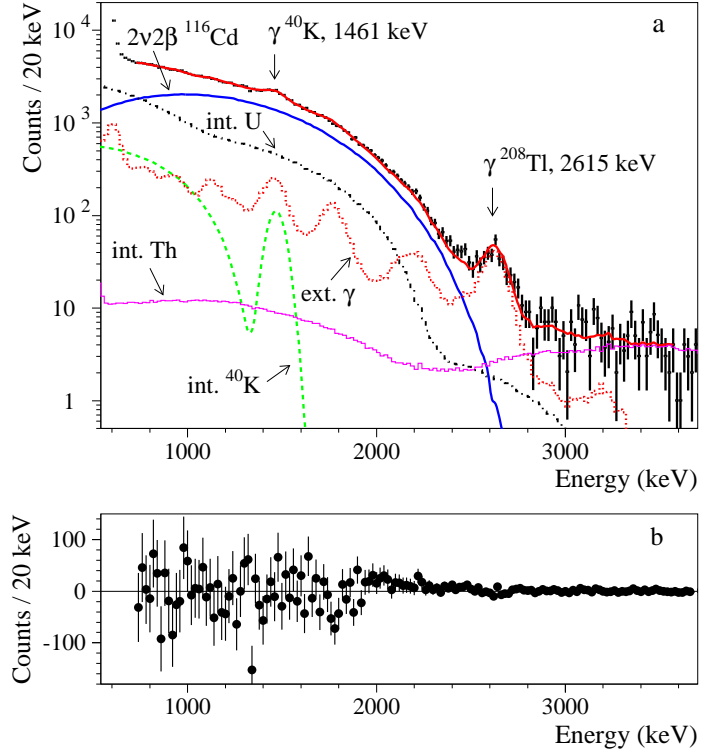}}
 \end{center}
 \vspace{-0.8cm}
\caption{The energy spectrum of $\gamma$($\beta$) events
accumulated over 26831 h with the $^{116}$CdWO$_4$ detectors
together with the main components of the background model: the
$2\nu2\beta$ decay of $^{116}$Cd (``$2\nu2\beta$''), internal
contaminations of the $^{116}$CdWO$_4$ crystals by U/Th, K (``int.
U'', ``int. Th'', ``int. $^{40}$K''), and contributions from
external $\gamma$ quanta (``ext. $\gamma$'') (a). The difference
between the experimental energy spectrum and the Monte Carlo
background model (b).}
 \label{fig:2n2b-sp}
\end{figure}

The energy spectrum of $\gamma$($\beta$) events selected by using
the PSD and front-edge analyzes was corrected taking into account
the efficiency of the simultaneous application of the PSD and
front-edge discrimination cuts presented in Fig.
\ref{fig:psd-fe-eff}. The corrected data accumulated over 26831~h
with the two $^{116}$CdWO$_4$ detectors are shown in Fig.
\ref{fig:2n2b-sp}. There is a clear signature of the $^{116}$Cd
$2\nu2\beta$ decay distribution in the data.

To estimate a half-life of $^{116}$Cd relatively to the
2$\nu$2$\beta$ decay, the spectrum was fitted by the background
model built from internal $^{40}$K, $^{90}$Sr, $^{90}$Y ($^{90}$Y
was assumed to be in equilibrium with $^{90}$Sr),
$^{110m}$Ag\footnote{Despite the long time after the crystal was
produced in the end of 2010, we cannot exclude presence of
cosmogenic nuclides (particularly of $^{110m}$Ag that was observed
in the crystals in the early measurements \cite{Barabash:2011})
since the scintillators were several times moved to surface for
treatment and the detector upgrade.}, beta active daughters of
$^{232}$Th and $^{238}$U, external gamma quanta from radioactive
contamination of the set-up by potassium, thorium and radium
(radioactive contamination of the copper shield, PMTs, and the
quartz light-guides were taken as free parameters), and the
$2\nu2\beta$ decay of $^{116}$Cd. All the models were simulated by
using the EGS4 simulation package \cite{EGS4}, the initial
kinematics of the particles emitted in the decays was given by an
event generator DECAY0 \cite{DECAY0}. The energy distribution of
the $2\nu2\beta$ decay of $^{116}$Cd (in total $5\times10^6$
decays were simulated in the both detectors) contains 98.86\% of
the simulated events. The loss of 1.14\% events is due to
rejection of escaped $\beta$ particles and bremsstrahlung $\gamma$
quanta by the liquid scintillator surrounding the $^{116}$CdWO$_4$
scintillators.

The experimental spectrum was fitted in the energy intervals
within $(640-1600)$ keV, for the starting point, and $(2800-3600)$
keV, for the final point, with a step 20 keV that gives the
${\chi}^{2}$/n.d.f. values (where n.d.f. is the number of degrees
of freedom) within $1.15-1.75$. The best fit
(${\chi}^{2}$/n.d.f.$~=142/124=1.15$) was achieved in the energy
interval $(720 - 3560)$ keV giving 126341 counts of the
$2\nu2\beta$ decay of $^{116}$Cd in the whole spectrum with a
statistical error 527 counts. The energy interval contains 73.55\%
of the whole $2\nu2\beta$ distribution, the effect to background
ratio is approximately 1.5. Using the number of $2\nu2\beta$
events, activity of $^{116}$Cd in the $^{116}$CdWO$_4$ crystals is
1.138(5) mBq/kg (only statistical error). The activity is
presented in Table \ref{tab:rc} together with activity of
$^{40}$K, and limits on activities of $^{110m}$Ag,
$^{90}$Sr$-^{90}$Y, $^{228}$Ra and $^{226}$Ra obtained from the
fits. Taking into account the number of $^{116}$Cd nuclei in the
crystal scintillators ($N=1.584\times10^{24}$), the half-life of
$^{116}$Cd relatively to the 2$\nu$2$\beta$ decay to the ground
state of $^{116}$Sn is (only statistical error):

\begin{center}
 $T_{1/2}$ = (2.630 $\pm$ 0.011)$\times$ 10$^{19}$ yr.\\
\end{center}

The main contribution to the systematic error comes from the
ambiguity of the background model, first of all, from the
uncertainty of radioactive contamination of the $^{116}$CdWO$_4$
crystals by $^{238}$U, since the $\beta$ spectrum of $^{234m}$Pa
(daughter of $^{238}$U) competes with the 2$\nu$2$\beta$ spectrum
of $^{116}$Cd (see Fig. \ref{fig:2n2b-sp} where the result of fit
and the main background components are shown). The estimations of
the contribution to the systematic error of uncertainties of the
internal radioactive contamination of the $^{116}$CdWO$_4$ crystal
scintillators and the external background from the details of the
set-up are given in Table \ref{tab:mod-er}. We assume that errors
of the internal radioactive contamination activities contribute to
the systematic error of the background model. If only limit on
activity is known (the case of $^{90}$Sr$-$$^{90}$Y, $^{110m}$Ag,
$^{228}$Ra, $^{226}$Ra) the number of counts $\pm$ error was taken
in the range from zero to the limit. Despite we cannot determine
exact activities of radioactive contamination in the set-up
details, the error of the external background model was taken from
the fit, since there are gamma peaks in the energy spectrum that
justify the radioactive contamination even if its exact
localization remains unknown.

\begin{table}[htb]
\caption{Contribution to the $T_{1/2}^{2\nu2\beta}$ systematic
error of the background model components due to internal
contamination of the $^{116}$CdWO$_4$ crystals and external
background. The number of counts in the experimental spectrum is
given too. The errors in the 3rd column are calculated in \% of
the $^{116}$Cd half-life.}
\begin{center}
\begin{tabular}{|l|l|l|}
 \hline
 Component of the                   & Number of counts          & Contribution  \\
 background model                   & in the energy interval    & to $T_{1/2}^{2\nu2\beta}$ error \\
  ~                                 & of fit $(720-3560)$ keV     & (\%) \\
 \hline
 Experimental data                  & 154956                    & - \\
 \hline
 $2\nu2\beta$                       & 92923                     & - \\
 \hline
 $^{40}$K                           & $6623\pm685$              & $\pm0.74$ \\
 \hline
 $^{90}$Sr$~-^{90}$Y                & $3_{-3}^{+1403}$          & $+1.51$ \\
  \hline
 $^{110m}$Ag                        & $170_{-170}^{+114}$       & $_{-0.18}^{+0.12}$ \\
  \hline
 $^{228}$Ac                         & $117_{-117}^{+173}$       & $_{-0.13}^{+0.19}$ \\
  \hline
 $^{228}$Th ($^{212}$Bi$+$$^{208}$Tl) & $714\pm55$              & $\pm0.06$ \\
  \hline
 $^{234m}$Pa                        & $33129\pm2455$            & $\pm2.64$ \\
  \hline
 $^{226}$Ra ($^{214}$Pb$+$$^{214}$Bi) & $500_{-500}^{+39}$        & $_{-0.54}^{+0.04}$ \\
 \hline
 $^{210}$Bi                         & $9244\pm550$              & $\pm0.59$ \\
 \hline
 Internal background model          & $50500_{-2663}^{+2969}$   & $_{-2.87}^{+3.19}$ \\
 \hline
 External background model          & $11388\pm557$             & $\pm0.60$ \\
 \hline
 Model of background (total)        & $61888_{-2721}^{+3021}$   & $_{-2.93}^{+3.25}$ \\
 \hline
\end{tabular}
\end{center}
\label{tab:mod-er}
\end{table}

To take into account imprecise knowledge of the set-up radioactive
contaminations localization, we have fitted the energy spectrum
presented in Fig. \ref{fig:2n2b-sp} by three ``extreme'' models
with radioactive contaminations localized in different details of
the set-up: 1) all the potassium, thorium and radium
contaminations are in the PMTs, 2) all are in the copper shield,
and 3) all are in the quartz light-guides. The extreme cases give
estimation of the systematic error due to the ambiguity of the
radioactive-contamination localization $_{-2.63}^{+1.54}$\% (see
Table \ref{tab:sys}). It should be stressed that the ``extreme''
fits are characterized by bigger values of ${\chi}^{2}$/n.d.f.,
that confirms our quite natural assumption that all the details of
the set-up (at least the ones, included in the background model)
have their own contamination. The variation of the $^{116}$Cd half-life
depending on the energy interval of fit was estimated as
$_{-1.02}^{+0.34}$\%. In fact, these errors are also related to
the uncertainty of the background model.

The error due to the detector energy scale instability is
estimated to be $\pm 1.72\%$. Then we assume that possible
uncertainties in the theoretical $2\nu2\beta$ decay spectral shape
contribute to the systematic error on the level of 1\%
\cite{Doi:1985}.

Finally, uncertainties of the PSD and front-edge analyzes cuts and
number of $^{116}$Cd nuclei contribute to the systematic error
too. All the systematic uncertainties of the $T_{1/2}$ are
summarized in Table \ref{tab:sys}.

\begin{table}[!b]
\caption{Systematic uncertainties of $T_{1/2}$ (\%).}
\begin{center}
\begin{tabular}{|l|l|}
 \hline
 Source                                     & Contribution \\
 \hline
 Number of $^{116}$Cd nuclei                & $\pm0.12$ \\
 \hline
 PSD and front-edge cuts efficiency         & $\pm1.2$ \\
 \hline
 Model of background                        & $_{-2.93}^{+3.25}$  \\
 \hline
 Localization of radioactive contaminations & $_{-2.63}^{+1.54}$ \\
 \hline
 Interval of the fit                        & $_{-1.02}^{+0.34}$ \\
 \hline
 Energy scale instability                   & $\pm 1.72$ \\
 \hline
 $2\nu2\beta$ spectral shape                & $\pm 1.0$ \\
 \hline
 Total systematic error                     & $_{-4.69}^{+4.30}$  \\
 \hline
 \end{tabular}
 \end{center}
 \label{tab:sys}
\end{table}

By summing all the systematic errors in square we obtain the
following half-life of $^{116}$Cd relatively to the $2\nu2\beta$
decay to the ground state of $^{116}$Sn:

\begin{center}
 $T_{1/2}=[2.630\pm 0.011(\mathrm{stat})^{+0.113}_{-0.123}(\mathrm{sys})]\times10^{19}$ yr.
\end{center}

Taking into account a comparatively small statistical error, the
final half-life value can be obtained by summing the errors in
quadrature:

\begin{center}
 $T_{1/2}=(2.63^{+0.11}_{-0.12})\times10^{19}$ yr.
\end{center}

\begin{figure}[htb!]
\begin{center}
\resizebox{0.54\textwidth}{!}{\includegraphics{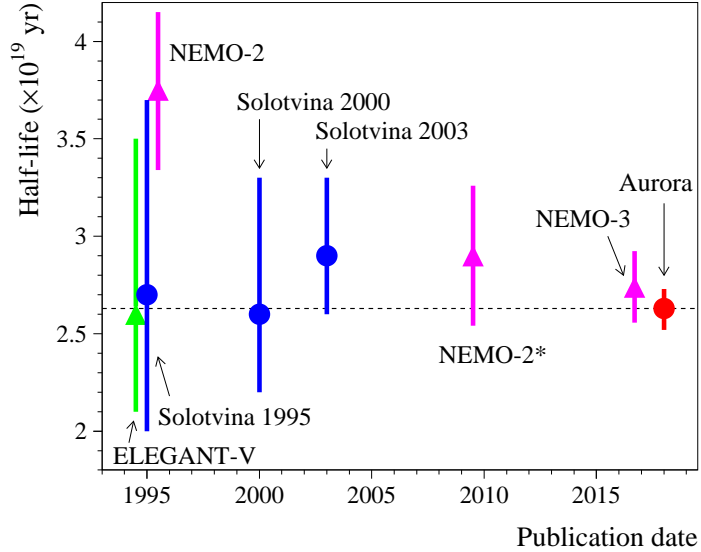}}
\end{center}
 \vspace{-0.8cm}
\caption{Comparison of the $^{116}$Cd 2$\nu$2$\beta$ half-life
obtained in the Aurora experiment with other experiments: ELEGANT
V \cite{Ejiri:1995}, Solotvina (three stages of the experiment
published in 1995, 2000 and 2003)
\cite{Danevich:2003b,Danevich:1995,Danevich:2000}, NEMO-2
\cite{NEMO2:1996}, and NEMO-3 \cite{NEMO3:2017}. A re-estimation
of the NEMO-2 experiment (NEMO-2*) \cite{NEMO2*:2010} is shown
too.}
\label{fig:08}
\end{figure}

The obtained half-life value is compared with the results of other
experiments in Table \ref{tab:2n2bres} and Fig. \ref{fig:08}.

\begin{table}[!b]
\caption{Effective nuclear matrix elements for $2\nu2\beta$ decay
of $^{116}$Cd to the ground state of $^{116}$Sn obtained by using
different calculations of the phase space factors.}
\begin{center}
\begin{tabular}{|l|l|l}
 \hline
 Phase space factor ($10^{-21}$ yr$^{-1}$), Reference   & Effective nuclear matrix element \\
 \hline
 2764 \cite{Kotila:2012}                               & $0.1173^{+0.0027}_{-0.0024}$ \\
 \hline
 3176 \cite{Kotila:2012} (SSD model)                   & $0.1094^{+0.0025}_{-0.0023}$ \\
 \hline
 2688 \cite{Mirea:2015}                                & $0.1189^{+0.0027}_{-0.0025}$ \\
 \hline
 \end{tabular}
 \end{center}
 \label{tab:NME}
\end{table}

The higher precision of the half-life value in the Aurora
experiment was achieved thanks to the certain advantages of the
radiopure, enriched in the isotope $^{116}$Cd scintillation
detectors: a high, accurately defined detection efficiency, in
contrast to the tracking experiments \cite{Ejiri:1995} and
\cite{NEMO2:1996} where detection efficiency depends on many
factors and typically cannot be estimated so precisely. In
addition, the $^{116}$Cd scintillation detectors used in the
present study have almost a twice higher energy resolution than
that in the Solotvina experiments (which also utilized enriched
$^{116}$CdWO$_4$ crystal scintillators, however of a lower quality
\cite{Danevich:2003b,Danevich:1995,Danevich:2000}). The higher
energy resolution, particularly to $\alpha$ particles, together
with the higher exposure of the experiment, allowed to estimate
the $^{238}$U activity in the $^{116}$CdWO$_4$ crystal
scintillators with a relative error $\approx 7\%$ (while only the
total alpha activity was estimated in the $^{116}$CdWO$_4$
scintillators used in the Solotvina experiments). The knowledge of
the activity, and therefore, activity of its daughter $\beta$
active $^{234m}$Pa (that competes with the 2$\nu$2$\beta$ spectrum
of $^{116}$Cd), allowed to reduce the model of background
uncertainty that is the main source of systematic error in the
scintillation experiments.

By using the half-life one can estimate an effective $NME^{eff}$
for the 2$\nu$2$\beta$ decay of $^{116}$Cd to the ground state of
$^{116}$Sn by using the following equation:
\begin{equation}
NME^{eff} = 1 / \sqrt{G_{2\nu2\beta} \times T_{1/2}},
\end{equation}

\noindent where $G_{2\nu2\beta}$ is the phase space factor.
Effective nuclear matrix elements calculated by using the space
factor from \cite{Kotila:2012,Mirea:2015} are presented in Table
\ref{tab:NME}.

\subsection{Search for 0$\nu$2$\beta$ decay of $^{116}$Cd}

There are no peculiarities in the experimental data which could be
ascribed to other possible $2\beta$ processes in $^{116}$Cd. A
lower limit on the half-life of $^{116}$Cd relatively to different
$2\beta$ decay channels and modes can be estimated by using the
following equation:

\begin{equation}
\lim T_{1/2} = N \cdot \eta \cdot t \cdot \ln 2 / \lim S,
\end{equation}

\noindent where $N$ is the number of $^{116}$Cd nuclei in the
$^{116}$CdWO$_4$ crystal scintillators, $\eta$ is the detection
efficiency for the process of decay, $t$ is the time of
measurements, and $\lim S$ is the number of events of the effect
searched for, which can be excluded at a given C.L.

To estimate a limit on the half-life of $^{116}$Cd relatively to
0$\nu2\beta$ decay to the ground state of $^{116}$Sn, we included
in the analysis also the data from the previous stage of the
experiment over 8493 h with a similar background counting rate of
$\approx0.1$ counts/(keV kg yr) in the energy interval $2.7-2.9$
MeV. Those data were not used for the analysis of the 2$\nu2\beta$
decay of $^{116}$Cd due to a much higher background counting rate
caused by rather high contamination of the Ultima Gold liquid
scintillator (surrounding the $^{116}$CdWO$_4$ crystal
scintillators) by potassium. The scintillator was replaced by the
radiopure one in the further stages of the experiment.

\begin{figure}[htb]
\begin{center}
\resizebox{0.54\textwidth}{!}{\includegraphics{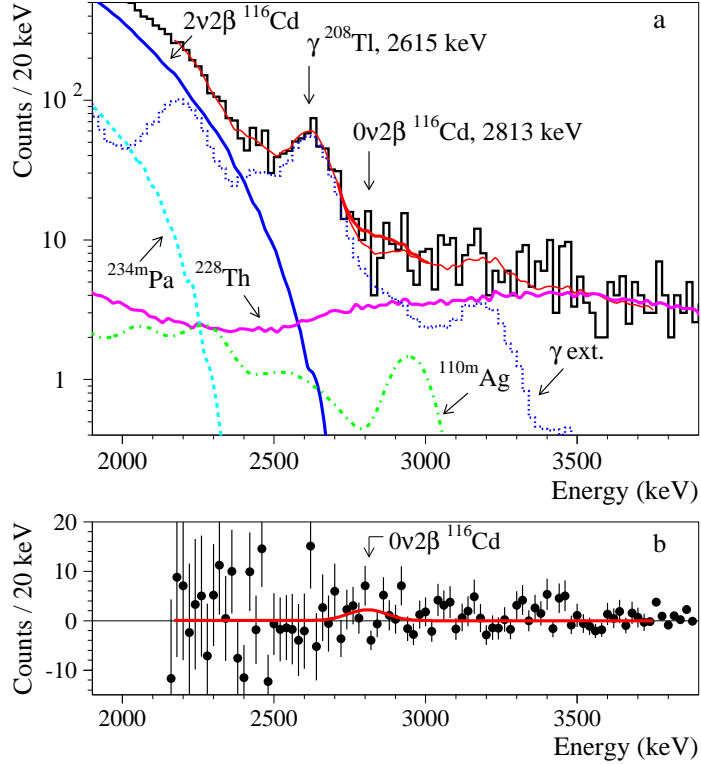}}
\end{center}
\caption{Part of the energy spectrum of $\gamma$($\beta$) events
accumulated over 35324 h with the $^{116}$CdWO$_4$ detectors
together with the background model: the $2\nu2\beta$ decay of
$^{116}$Cd, the internal contamination of the $^{116}$CdWO$_4$
crystals by $^{110m}$Ag, $^{228}$Th and $^{234m}$Pa, and the
contribution from external $\gamma$ quanta (``$\gamma$ ext.''). A
peak of the $0\nu2\beta$ decay of $^{116}$Cd excluded at 90\% C.L.
is shown too (a). The difference between the experimental energy
spectrum and the Monte Carlo background model (points with error
bars) together with the excluded peak of the $0\nu2\beta$ decay of
$^{116}$Cd (solid line) (b).}
 \label{fig:0n2b}
\end{figure}

The sum energy spectrum over 35324 h with the background counting
rate 0.146(12) counts/(keV yr kg), corrected for the efficiency of
the PSD analysis (Fig. \ref{fig:psd-fe-eff}, c), is presented in
Fig. \ref{fig:0n2b}. The spectrum was approximated in the energy
intervals $(2.1-2.3)$ MeV -- $(3.3-3.7)$ MeV with a step 20 keV by
the background model constructed from the distributions of the
$0\nu2\beta$ decay (an effect searched for), $2\nu2\beta$ decay of
$^{116}$Cd with the half-life $2.63_{-0.12}^{+0.11}\times10^{19}$
yr, the internal contamination of the crystals by $^{110m}$Ag,
$^{228}$Th and $^{234m}$Pa (bounded within the values or limits
presented in Table \ref{tab:rc}), and the contribution from
external $\gamma$ quanta from contamination of the set-up by
radium (only $^{214}$Bi was considered due to the large enough
energy of $\beta$ decay) and thorium ($^{208}$Tl). The best fit
(${\chi}^{2}$/n.d.f.$~=70.6/70=1.01$) achieved in the energy
interval $2160-3740$ keV gives an area of the peak searched for
$S=-4.5\pm 14.2$ counts, that is no evidence of the effect. It
should be stressed that the fit of the peak area (i.e. $S=-4.5\pm
14.2$ counts) includes only statistical errors coming from the
data fluctuations, and that systematic contributions have not been
included in the quoted value of the peak area error obtained with
90\% C.L. In accordance with \cite{Feldman:1998}, we took $\lim
S=19.1$ counts that can be excluded at 90\% C.L. Taking into
account the detection efficiency $\eta=0.9597$ (the part of
simulated events remaining in the whole energy distribution due to
escape of $\beta$ particles and bremsstrahlung $\gamma$ quanta),
the new limit on the $0\nu2\beta$ decay of $^{116}$Cd to the
ground state of $^{116}$Sn is set as:

\begin{center}
$T_{1/2}\geq 2.2 \times 10^{23}$ yr at 90\% C.L.
\end{center}

Similar estimations can be obtained for the experimental
sensitivity, by using simple consideration of the background
statistics in the region of interest. The total number of events
in the energy interval $(2720-2920)$ keV (where 81.43\% of the
peak is concentrated) is 113 counts that leads to a $\lim S=20.3$
counts at 90\% C.L. according to the procedure proposed by Feldman
and Cousins for an expected background and no true signal (Table
XII in \cite{Feldman:1998}). The approach provides a half-life
limit $T_{1/2}\geq 1.7 \times 10^{23}$ yr. Another estimation of
$\lim S=17.4$ counts can be obtained as $1.64\times
\sqrt{N_{BG}}$, where $N_{BG}=113$ is number of background counts
in the energy interval $(2720-2920)$ keV. It corresponds to the
half-life limit $T_{1/2}\geq 2.0 \times 10^{23}$ yr, that again is
near to the result obtained from the fit.

Assuming the mass mechanism of $0\nu2\beta$ decay with light
neutrino exchange, we can estimate a limit on the effective
Majorana neutrino mass $\langle m_{\nu}\rangle$ by using the
following equation for the $0\nu2\beta$ decay rate:

\begin{equation}
[T^{0\nu}_{1/2}]^{-1} = g^4_{A}G^{0\nu}|M^{0\nu}|^2 \frac{\langle
m_\nu\rangle^2}{m_e^2},
\end{equation}

\noindent where $g_A$ is the axial vector coupling constant,
$G^{0\nu}$ is the phase space factor that depends on $Z$ and the
nuclear transition energy $Q_{2\beta}$, $M^{0\nu}$ is the NME for
$0\nu2\beta$ decay, $m_e$ is the electron mass. In our analysis we
use the phase space factor from \cite{Kotila:2012} and the axial
vector coupling constant $g_A=1.27$.  By using the recent
$M^{0\nu}$ obtained in the framework of the density functional
theory based on a non-relativistic \cite{Vaquero:2013} and a
relativistic \cite{Song:2017} energy density functional theory,
the quasiparticle random-phase approximation \cite{Simkovic:2013},
the proton-neutron quasiparticle random-phase approximation
\cite{Hyvarinen:2015}, and the microscopic interacting boson model
\cite{Barea:2015} we have obtained the following interval of the
effective Majorana neutrino mass limits:

\begin{center}
 $\langle$$m_\nu$$\rangle \leq (1.0-1.7)$ eV at 90\% C.L.
\end{center}

\begin{figure}[htb]
\begin{center}
\resizebox{0.54\textwidth}{!}{\includegraphics{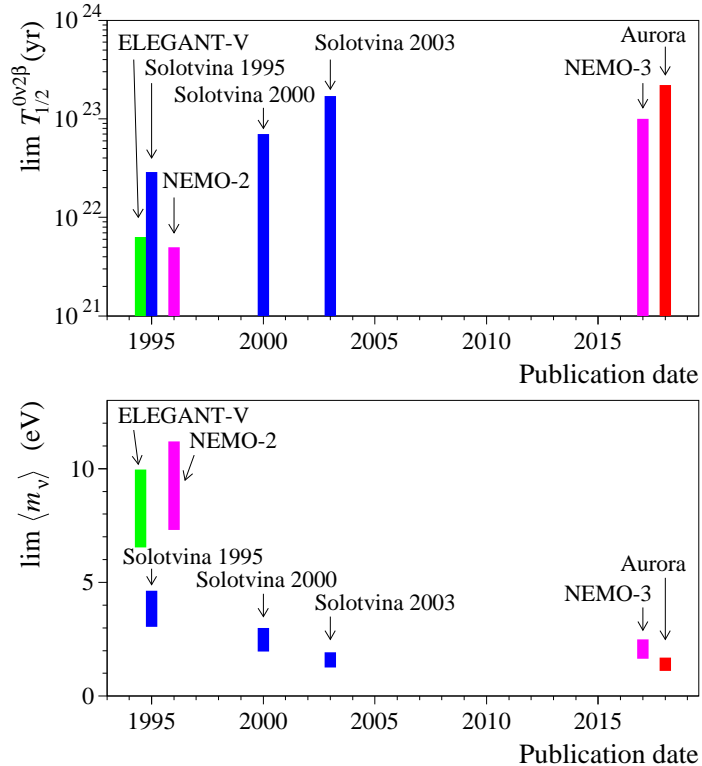}}
\end{center}
 \vspace{-0.8cm}
\caption{Comparison of the $^{116}$Cd $0\nu$2$\beta$ half-life
limits obtained in the Aurora experiment with other experiments:
ELEGANT V \cite{Ejiri:1995}, Solotvina (three stages of the
experiment published in 1995, 2000 and 2003)
\cite{Danevich:2003b,Danevich:1995,Danevich:2000}, NEMO-2
\cite{NEMO2:1996}, and NEMO-3 \cite{NEMO3:2017} (upper panel).
Comparison of the effective Majorana neutrino mass limits obtained
in the Aurora experiment with estimations of
$\langle$$m_\nu$$\rangle$ limits obtained in the other
experiments. The intervals of the effective Majorana neutrino mass
limits were calculated by using the phase space factor from
\cite{Kotila:2012}, the axial vector coupling constant $g_A=1.27$,
and the same $M^{0\nu}$ that have been utilized to estimate the
neutrino-mass limits interval in the present experiment
\cite{Barea:2015,Simkovic:2013,Vaquero:2013,Hyvarinen:2015,Song:2017}
(lower panel).}
 \label{fig:nu_m}
\end{figure}

The obtained limits on half-life and on the effective Majorana neutrino 
mass are compared with the limits of other experiments in Fig. \ref{fig:nu_m}. 
Neutrinoless $2\beta$ decay can be mediated by different
mechanisms, particularity by hypothetical right-handed currents
admixture in the weak interaction. The following limits were set
on the parameters of the admixtures using calculations
\cite{NEMO2:1996,Staudt:1990,Pantis:1996}: $\langle \eta \rangle
\leq (1.6-21)\times 10^{-8}$ and $\langle \lambda \rangle \leq
(1.8-22) \times 10^{-6}$. In accordance with \cite{R-parity} the
value of the coupling constant $\lambda'_{111}$ in the R-parity
violating minimal supersymmetric standard model is restricted by
the $T_{1/2}^{0\nu}$ limit to $\lambda'_{111} \leq
2.5\times10^{-4} \times f$ at 90\% C.L., where $f = (m_{\tilde{q}}
/ 100 \textrm{ GeV})^2 \times (m_{\tilde{g}} / 100 \textrm{
GeV})^{1/2}$; $m_{\tilde{q}}$ and $m_{\tilde{g}}$ are the squark
and gluino masses. Also an interval of lower limits on the heavy
neutrino mass was estimated assuming the $0\nu2\beta$ decay
mechanism of exchanging by heavy Majorana neutrino. By using the
nuclear matrix elements ($M^{0\nu_h}=110-302$) calculated in
\cite{Barea:2015,Hyvarinen:2015,Song:2017,Faessler:2014}, the
phase-space factor ($G_{0\nu} = 16.7\times 10^{-15}$ yr$^{-1}$)
from \cite{Kotila:2012}, and $g_A = 1.27$ the mass of heavy
Majorana neutrino is restricted as $|\langle
m_{\nu_{h}}^{-1}\rangle |^{-1} \geq (10-28)\times10^6$ GeV.

\subsection{Search for $2\beta$ transitions to excited levels of $^{116}$Sn}

The $2\beta$ decay can also proceed through transitions to excited
levels of the daughter nucleus. Studies of the latter transitions
allow to extract supplementary information about the $2\beta$
process. Up to now $2\nu2\beta$ decay to the first $0^+$ excited
\begin{figure}[b!]
 \begin{center}
 \resizebox{0.54\textwidth}{!}{\includegraphics{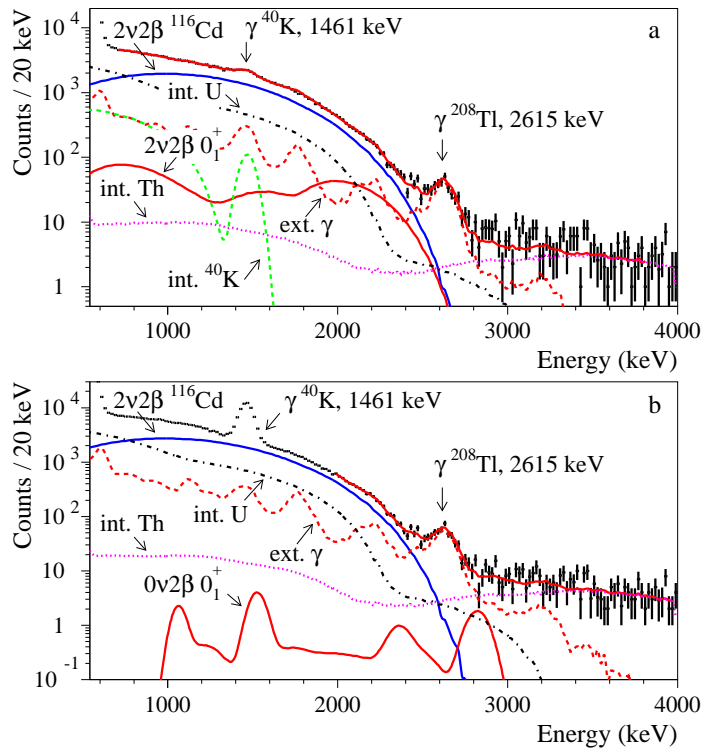}}
 \end{center}
 \vspace{-0.8cm}
 \caption{The energy spectra of $\gamma$($\beta$) events
measured by the $^{116}$CdWO$_4$ detectors over 26831 h (a) and
35324 h (b), corrected on the efficiency of the PSD and front-edge
cuts, together with the main components of the background model:
the g.s. to g.s. $2\nu2\beta$ decay of $^{116}$Cd, internal
contaminations of the $^{116}$CdWO$_4$ crystals by U/Th, K ("int.
U", "int. Th", "int. $^{40}$K"), and contributions from external
$\gamma$ quanta ("ext. $\gamma$"). The fit and the excluded
distributions of the $2\nu2\beta$ (a) and $0\nu2\beta$ (b) decay
of $^{116}$Cd to the first excited $0^+$ 1757 keV level of
$^{116}$Sn are shown too.}
 \label{fig:13}
\end{figure}
state of daughter nuclei was detected in $^{100}$Mo and $^{150}$Nd
(see review \cite{Barabash:2017}). The theoretical predictions for
such transition in $^{116}$Cd are on the level of
$T_{1/2}\sim10^{21}-10^{24}$ yr (see review \cite{Barabash:2017}).
So, there is a chance to detect this transition in $^{116}$Cd too.
As it was noted in \cite{Simkovic:2002}, the detection of
$0\nu2\beta$ transition to excited levels would give an additional
possibility to distinguish mechanisms of the $0\nu2\beta$ decay if
observed.

We set limits on $2\beta$ transition to several lowest excited
levels of $^{116}$Sn by fit of the data in different energy
intervals. For instance, the energy spectrum measured by the
$^{116}$CdWO$_4$ detectors over 26831 h was fitted in the energy
intervals from $(700-1300)$ keV to $(3200-4000)$ keV by the model
similar to the one used for the $2\nu2\beta$ decay of $^{116}$Cd
to the ground state of $^{116}$Sn with added simulated
distribution of $2\nu2\beta$ decay of $^{116}$Cd to the first
excited $0^+$ 1757 keV level of $^{116}$Sn. The best fit, achieved
in the energy interval $(700-3800)$ keV ($\chi^2$/n.d.f.~=~1.13),
provides area of the effect $2111\pm1807$ counts, that gives no
evidence for the effect searched for. The $\lim S=5075$ counts (at
90\% C.L.) can be obtained assuming Gaussian errors
\cite{Feldman:1998}. Taking into account the detection efficiency
88.94\% we get a half-life limit $T_{1/2}\ge 5.9\times 10^{20}$
yr. The excluded distribution of the $2\nu2\beta$ decay of
$^{116}$Cd to the first excited $0^+$ 1757 keV level of $^{116}$Sn
is shown in Fig. \ref{fig:13} (a).

\begin{table*}[!b]
\caption{Summary of the obtained results on 2$\beta$ processes in
$^{116}$Cd. The limits are given at 90\% C.L., except of the
results of \cite{Barabash:1990}, obtained at 68\% C.L.}
    \begin{center}

 \begin{tabular}{llll}
 \hline
 Decay mode             & Transition,           & $T_{1/2}$ (yr)                    & Best previous \\
 ~                      & level of $^{116}$Sn   &   ~                               & limits (yr) \\
  ~                     & (keV)                 &   ~                               & Reference \\
 \hline
$2\nu$                  & g.s.                  & $(2.63_{-0.12}^{+0.11})\times10^{19}$ yr   & see Table \ref{tab:2n2bres} and Fig. \ref{fig:08} \\
$2\nu$                  & $2^{+}$ (1294)        & $\geq 9.8\times10^{20}$           & $\geq 2.3\times10^{21}$ \cite{Piepke:1994} \\
$2\nu$                  & $0^{+}$ (1757)        & $\geq 5.9\times10^{20}$           & $\geq 2.0\times10^{21}$ \cite{Piepke:1994} \\
$2\nu$                  & $0^{+}$ (2027)        & $\geq 1.1\times10^{21}$           & $\geq 2.0\times10^{21}$ \cite{Piepke:1994} \\
$2\nu$                  & $2^{+}$ (2112)        & $\geq 2.5\times10^{21}$           & $\geq 1.7\times10^{20}$ \cite{Barabash:1990} \\
$2\nu$                  & $2^{+}$ (2225)        & $\geq 7.5\times10^{21}$           & $\geq 1.0\times10^{20}$ \cite{Barabash:1990} \\
$0\nu$                  & g.s.                  & $\geq 2.2\times10^{23}$           & $\geq 1.7\times10^{23}$ \cite{Danevich:2003b} \\
$0\nu$                  & $2^{+}$ (1294)        & $\geq 7.1\times10^{22}$           & $\geq 2.9\times10^{22}$ \cite{Danevich:2003b} \\
$0\nu$                  & $0^{+}$ (1757)        & $\geq 4.5\times10^{22}$           & $\geq 1.4\times10^{22}$ \cite{Danevich:2003b} \\
$0\nu$                  & $0^{+}$ (2027)        & $\geq 3.1\times10^{22}$           & $\geq 0.6\times10^{22}$ \cite{Danevich:2003b} \\
$0\nu$                  & $2^{+}$ (2112)        & $\geq 3.7\times10^{22}$           & $\geq 1.7\times10^{20}$ \cite{Barabash:1990} \\
$0\nu$                  & $2^{+}$ (2225)        & $\geq 3.4\times10^{22}$           & $\geq 1.0\times10^{20}$ \cite{Barabash:1990} \\
$0\nu\chi^0~n=1$        & g.s.                  & $\geq 8.2\times10^{21}$           & $\geq 8.5\times10^{21}$ \cite{NEMO3:2017} \\
$0\nu\chi^0~n=2$        & g.s.                  & $\geq 4.1\times10^{21}$           & $\geq 1.7\times10^{21}$ \cite{Danevich:2003b} \\
$0\nu\chi^0~n=3$        & g.s.                  & $\geq 2.6\times10^{21}$           & $\geq 0.8\times10^{21}$ \cite{Danevich:2003b} \\
$0\nu\chi^0\chi^0~n=3$  & g.s.                  & $\geq 2.6\times10^{21}$           & $\geq 0.8\times10^{21}$ \cite{Danevich:2003b} \\
$2\nu LV~n=4$           & g.s.                  & $\geq 1.2\times10^{21}$           & -- \\
$0\nu\chi^0\chi^0~n=7$  & g.s.                  & $\geq 8.9\times10^{20}$           & $\geq 4.1\times10^{19}$ \cite{Arnold:2000}  \\
 \hline
 \end{tabular}
 \end{center}
  \label{tab:res}
\end{table*}

For the $0\nu2\beta$ decay of $^{116}$Cd to the first excited
$0^+$ 1757 keV level of $^{116}$Sn, the highest sensitivity was
achieved by analysis of the data recorded with the
$^{116}$CdWO$_4$ detectors over 35324 h. The spectrum, see Fig.
\ref{fig:13} (b), was fitted in the energy intervals from
$(1500-2000)$ keV to $(3200-4000)$ keV by the same model, however,
without contribution from internal and external $^{40}$K. In this
case the best fit was achieved in the energy interval $1980-3900$
keV ($\chi^2$/n.d.f.~$=0.964$) with the effect area $-7\pm57$
counts that again gives no evidence of the effect observation. An
estimation of $\lim S=87$ counts (90\% C.L.) was obtained by using
the Feldman-Cousins recommendations \cite{Feldman:1998}. The
detection efficiency for the neutrinoless transition is 88.23\%,
that leads to the half-life limit $T_{1/2}\ge 4.5\times 10^{22}$
yr. Limits on other $2\beta$ transitions of $^{116}$Cd to excited
levels of $^{116}$Sn were obtained in a similar way. They are
presented in Table \ref{tab:res}, where results of the most
sensitive previous experiments are given for comparison.

\subsection{Search for 2$\beta$ decay with majoron emission and Lorentz violation}

Spontaneous violation of global $B-L$ symmetry in gauge theories
leads to the existence of a massless Goldstone boson, the majoron
($\chi^0$). The majoron, if it exists, could play a significant
role in the history of the early Universe and in the evolution of
stars. In addition, majoron could play the role of the dark matter
particle (see, for example, \cite{Lattanzi:2013,Kazanas:2004}). In
the original majoron models, the majoron is part of an electroweak
singlet \cite{Chikashige:1980,Chikashige:1981}, doublet
\cite{Aulakh:1982}, or triplet \cite{Gelmini:1981}. The models of
a doublet and triplet majoron were disproved in 1989 by the data
on the decay width of the Z$^0$ boson that were obtained at the
LEP \cite{Caso:1998}. Despite this, some new models were proposed
\cite{Mohapatra:1991,Berezhiani:1992}, where $0\nu\chi^0$ $2\beta$
decay is possible and where there are no contradictions with the
LEP data. A $2\beta$ decay model that involves the emission of two
majorons was proposed within Supersymmetric theories
\cite{Mohapatra:1988}, and several other models of the majoron
were proposed in the 1990s. By the term ``majoron'' one means
massless or light bosons that are associated with neutrinos. In
these models, the majoron can carry a lepton charge and is not
required to be a Goldstone boson \cite{Burgess:1993,Burgess:1994}.
A decay process that involves the emission of two majorons is also
possible \cite{Mohapatra:1988,Bamert:1995}. In models featuring a
vector majoron, the majoron is the longitudinal component of a
massive gauge boson emitted in $2\beta$ decay \cite{Carone:1993}.
In the work \cite{Mohapatra:2000} a ``bulk'' majoron model was
proposed in the context of the ``brane-bulk'' scenario for
particle physics. Classification of majoron models (related to
$2\beta$ decay) can be found in \cite{Arnold:2000}. The shape of
the two-electron energy sum distribution depends on the ``spectral
index'' $n$ defined by the phase space of the emitted particles $G
\sim(Q_{2\beta}-T)^n$, where $Q_{2\beta}$ is the energy released
in the decay and $T$ is the energy of the two electrons (the
ordinary $2\nu2\beta$ decay has the spectral index $n=5$). The
single majoron decay $2\beta\chi^0$ is possible with $n=1$, 2 and
3. The models for the emission of two majorons
$2\beta\chi^0\chi^0$ correspond to $n=3$ and 7. The half-life for
ordinary majoron with spectral index $n=1$ can be written as:

\begin{equation}
[T_{1/2}^{0\nu\chi^0}]^{-1} = G_{0\nu\chi^0} \cdot g_A^4 \cdot
\langle g_{ee}\rangle^2 \cdot \mid{ M_{0\nu \chi^0}}\mid^2,
\end{equation}

\noindent where $G_{0\nu\chi^0}$ is the phase space factor (which
is accurately known \cite{Kotila:2015}), $M_{0\nu \chi^0}$ is the
nuclear matrix element (the same as for $0\nu\beta\beta$ decay),
$\langle g_{ee}\rangle$ is the coupling constant of the majoron to
the neutrino, and $g_A$ is the axial-vector coupling constant.

In decay with emission of two majorons, we have:

\begin{equation}
[T_{1/2}^{0\nu\chi^0 \chi^0}]^{-1} = G_{0 \nu \chi^0 \chi^0} \cdot
g_A^4 \cdot \langle g_{ee}\rangle^4 \cdot \mid{ M_{0\nu \chi^0
\chi^0}}\mid^2.
\end{equation}

The Lorentz invariance (LI) is one of the founding principles of
modern physics, but it could be only approximate symmetry of our
local space-time possibly modified at some scale outside of our
experience. As any fundamental principle, LI should be checked
with the highest available to-date sensitivity (see, for example,
reviews \cite{Kostelecky:2011,Tasson:2014}). As it was noted in
\cite{Diaz:2013,Diaz:2014}, LI could be tested also in $2\beta$
decay experiments as LI violation leads to energy spectra of
emitted particles different from those in usual $2\nu2\beta$
process. This alteration of the electron-sum spectrum in the
$2\nu2\beta$ decay has been explored by the EXO-200 experiment,
obtaining the first experimental limit on the relevant coefficient
for the Lorentz violation (LV) \cite{Albert:2016}. In addition,
CPT-violating Majorana couplings in the Standard-Model extensions
can trigger $0\nu2\beta$ decay even for a negligible Majorana mass
\cite{Diaz:2014}.

\begin{figure}[htb!]
\begin{center}
\resizebox{0.54\textwidth}{!}{\includegraphics{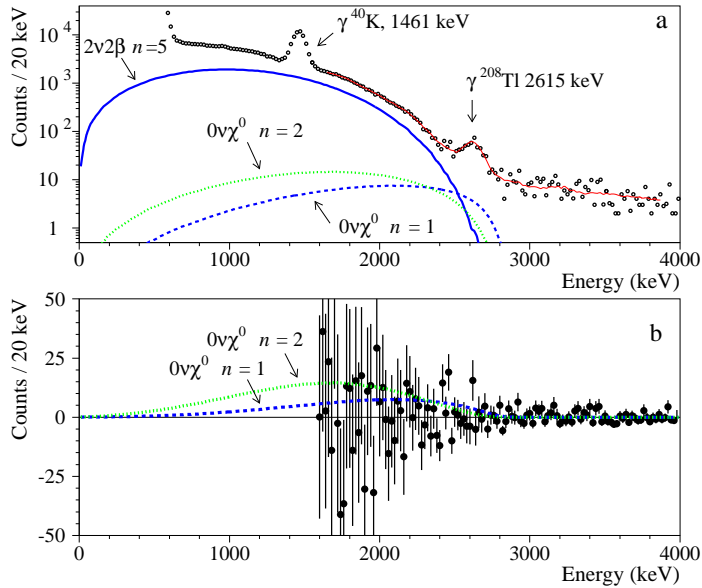}}
\end{center}
\caption{Energy spectrum of $^{116}$CdWO$_4$ detectors acquired
over 35324 h. The fit of the data, the $2\nu2\beta$ spectrum of
$^{116}$Cd and excluded at 90\% C.L. distributions for
neutrinoless double-beta decay of $^{116}$Cd with majorons
emission ($n=1$ and $n=2$) are shown (a). Difference between the
experimental data and the background model together with the
excluded distributions (b).}
 \label{fig:majorons12}
\end{figure}

\begin{figure}[htb!]
\begin{center}
\resizebox{0.54\textwidth}{!}{\includegraphics{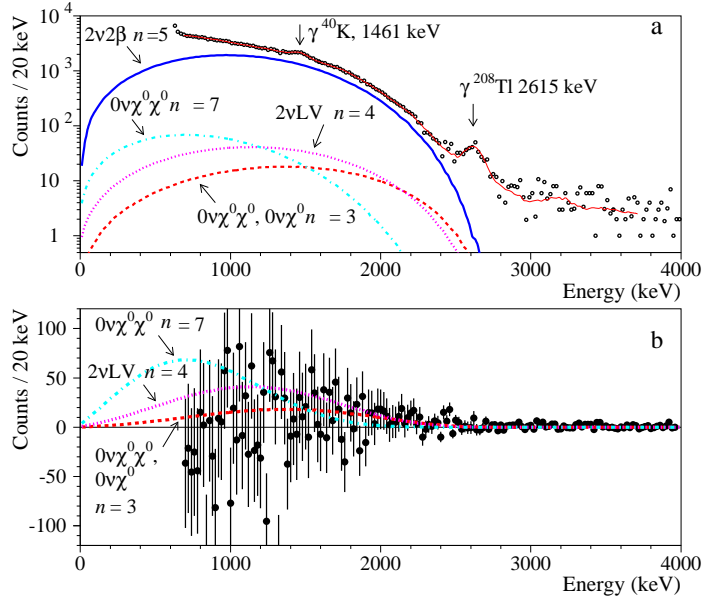}}
\end{center}
\caption{Energy spectrum of $^{116}$CdWO$_4$ detectors acquired
over 26831 h. The fit of the data, the $2\nu2\beta$ spectrum of
$^{116}$Cd and excluded at 90\% C.L. distributions for
neutrinoless double-beta decay of $^{116}$Cd with majorons
emission ($n=3$ and $n=7$), and for Lorentz-violating ($n=4$)
$2\nu2\beta$ decay of $^{116}$Cd are shown (a). Difference between
the experimental data and the background model together with the
excluded distributions (b).} \label{fig:majorons347}
\end{figure}

Search for the $0\nu2\beta$ decay with majorons emission and
$2\nu2\beta$ Lorentz-violating decay was realized by using an
approach similar to the utilized for the investigations of the
$2\beta$ decay to the excited levels of $^{116}$Sn. For instance
the experimental energy spectrum gathered over 35324 h was
analyzed to set a limit on the $0\nu2\beta$ decay with single
majoron emission ($n=1$). The fit in the energy interval
$(2200-3860)$ keV ($\chi^2$/n.d.f.=1.13) gives an area of the
simulated distribution $113\pm241$ counts that corresponds to
$\lim S=533$ counts (the fit and excluded $0\nu\chi^0$
distribution are shown in Fig. \ref{fig:majorons12}, a). Taking
into account the detection efficiency of the decay (98.38\%) the
half-life limit can be set as
$T_{1/2}^{0\nu\chi^0}(n=1)\geq8.2\times 10^{21}$ yr at 90\% C.L.
Limits on other possible neutrinoless double-beta processes with
majorons emission and the Lorentz-violating $2\nu2\beta$ decay
were set in a similar way (see Fig. \ref{fig:majorons347}). All
the results of the experiment are summarized in Table
\ref{tab:res}.

Using the limit on the $0\nu2\beta$ decay with majoron emission
with $n=1$, the phase space integral calculations
\cite{Kotila:2015}, and the axial vector coupling constant
$g_A=1.27$, we get an upper limit on the coupling constant with
the majoron emission $\langle g_{ee}\rangle \leq
(6.1-9.3)\times10^{-5}$.

To derive the limits on $\langle g_{ee} \rangle$ in other models
with one or two majoron(s) emissions and $n=3,7$, we used the
nuclear matrix elements and the phase space factors calculated in
\cite{Hirsch:1996}. The results are given in Table \ref{tab:par}.

In the Lorentz-violated $2\nu2\beta$ decay \cite{Diaz:2014}, the
differential decay rate is described by expression

\begin{equation}
d\Gamma/dt_1dt_2 = C \cdot e_1p_1F(t_1,Z) \cdot e_2p_2F(t_2,Z)
\cdot [(t_0-t_1-t_2)^5 +
10\mathring{a} ^{(3)}_{\mathrm{of}}(t_0-t_1-t_2)^4],
\end{equation}

\noindent where $C$ is the normalizing constant, $t_i$ is the
kinetic energy of the $i$-th electron (all energies here are in
units of the electron mass $m_ec^2$), $e_i=t_i+1$ is the total
energy of $i$-th particle, $p_i$ is its momentum
$p_i=\sqrt{t_i(t_i+2)}$ (in units of $m_ec$), $t_0$ is the energy
release, and $F(t,Z)$ is the Fermi function which takes into
account the influence of the electric field of the nucleus on the
emitted electrons ($Z$ is atomic number of the daughter nucleus).
Thus, the shape and the total rate in the LV-$2\nu2\beta$ decay
are different in comparison to the usual $2\nu2\beta$ process. The
total rate is:

\begin{equation}
\Gamma = \Gamma_{2\nu} + \Gamma_{2\nu \mathrm{LV}},
\end{equation}

\noindent where

\begin{equation}
\Gamma_{2\nu} = C I_5, ~~~~~ \Gamma_{2\nu \mathrm{LV}} =
10\mathring{a} ^{(3)}_{\mathrm{of}} \cdot C I_4,
\end{equation}

\begin{equation}
I_5 = \int_0^{t_0} dt_1 e_1p_1F(t_1,Z) \int_0^{t_0-t_1} dt_2
e_2p_2F(t_2,Z) (t_0-t_1-t_2)^5,
\end{equation}

\begin{equation}
I_4 = \int_0^{t_0} dt_1 e_1p_1F(t_1,Z) \int_0^{t_0-t_1} dt_2
e_2p_2F(t_2,Z) (t_0-t_1-t_2)^4.
\end{equation}

The LV amplitude (or its limit) can be find as

\begin{equation}
10\mathring{a} ^{(3)}_{\mathrm{of}} = \frac{\Gamma_{2\nu
\mathrm{LV}}}{\Gamma_{2\nu}} \cdot \frac{I_5}{I_4} =
\frac{T_{1/2}^{2\nu}}{T_{1/2}^{2\nu \mathrm{LV}}} \cdot
\frac{I_5}{I_4}.
\end{equation}

The $I_4$, $I_5$ integrals can be calculated numerically using
tabulated values of the Fermi function \cite{Beh69}\footnote{Using
the Primakoff-Rosen approximation \cite{Pri59} $F(t,Z) \sim e/p$
(which works well for $\beta^-$ and $2\beta^-$ decays), it is
possible to calculate the integrals analytically: $I_5 =
t_0^7(t_0^4+22t_0^3+220t_0^2+990t_0+1980)/83160$, $I_4 =
t_0^6(t_0^4+20t_0^3+180t_0^2+360t_0+1260)/37800$. This gives:
$\mathring{a} ^{(3)}_{\mathrm{of}} = (T_{1/2}^{2\nu} /
T_{1/2}^{2\nu \mathrm{LV}}) \times 4.55 \times 10^{-2} \times
Q_{2\beta} \cdot (t_0^4+22t_0^3+220t_0^2+990t_0+1980) /
(t_0^4+20t_0^3+180t_0^2+360t_0+1260)$. }. With the values obtained
in this work: $T_{1/2}^{2\nu} = 2.63\times10^{19}$ yr and $\lim
T_{1/2}^{2\nu \mathrm{LV}} = 1.2\times10^{21}$ yr, we get limit
$\mathring{a} ^{(3)}_{\mathrm{of}} \leq 4.0\times10^{-6}$ GeV, at
the same level as that obtained in the EXO-200 experiment for
$^{136}$Xe.

A summary on limits on lepton-number violating parameters obtained
in the present work is given in Table \ref{tab:par}.

\begin{table}[htb]
\caption{Limits on lepton-number violating parameters. The limits
are given at 90\% C.L.}
\begin{center}
\begin{tabular}{|l|l|}
 \hline
 Parameter                                                                   & Limit \\
 \hline
 Effective light Majorana neutrino mass $\langle m_{\nu} \rangle$           & $\leq(1.0-1.7)$ eV \\
 \hline
 Effective heavy Majorana neutrino mass $|\langle m_{\nu_{h}}^{-1}\rangle |^{-1}$  & $ \geq (10-28)\times10^6$ GeV \\
 \hline
 Right-handed current admixture $\langle \lambda \rangle$                   & $\leq (1.8-22)\times10^{-6}$ \\
 \hline
 Right-handed current admixture $\langle \eta \rangle$                      & $\leq (1.6-21)\times10^{-8}$ \\
 \hline
 Coupling constant of neutrino with majoron $\langle g_{ee} \rangle$ & \\
 $\chi^0,~n=1$                                                              & $\leq(6.1-9.3)\times10^{-5}$ \\
 $\chi^0,~n=3$                                                              & $\leq 7.7 \times10^{-2}$ \\
 $\chi^0\chi^0,~n=3$                                                        & $\leq (0.69-6.9)$ \\
 $\chi^0\chi^0,~n=7$                                                        & $\leq (0.57-5.7)$ \\
 \hline
 R-parity violating parameter $\lambda'_{111}$                              & $\leq 2.5 \times 10^{-4} \times f$ (see text) \\
 \hline
 Lorentz-violating parameter $\mathring{a} ^{(3)}_{\mathrm{of}}$            & $\leq 4.0\times10^{-6}$ GeV \\
 \hline

 \end{tabular}
 \end{center}
 \label{tab:par}
\end{table}

\section{CONCLUSIONS}

The Aurora experiment to investigate 2$\beta$ processes in
$^{116}$Cd with 1.162~kg of enriched $^{116}$CdWO$_4$
scintillators is finished after about 5 years of data taking at
the Gran Sasso underground laboratory of I.N.F.N. (Italy). The
half-life of $^{116}$Cd relatively to the $2\nu2\beta$ decay to
the ground state of $^{116}$Sn is measured with the highest
up-to-date accuracy: $T_{1/2}=(2.63^{+0.11}_{-0.12})\times10^{19}$
yr. The statistical error of the value is negligible (0.4\%),
while the main sources of the systematic error are the
uncertainties of the background model and of the detector energy
scale, and the pulse-shape discrimination cuts efficiency.
Two-neutrino and neutrinoless $2\beta$ transitions of $^{116}$Cd
to several excited levels of $^{116}$Sn are restricted at the
level of $T_{1/2}>10^{20}-10^{22}$ yr.

A new half-life limit on the $0\nu2\beta$  decay of $^{116}$Cd to
the ground state of $^{116}$Sn is set as $T_{1/2}$ $\ge$
2.2$\times$ 10$^{23}$ yr at 90\% C.L., that corresponds to the
effective Majorana neutrino mass limits $\langle$$m_\nu$$\rangle$
$\leq$ $(1.0-1.7)$ eV, depending on the nuclear matrix elements
used in the analysis. Neutrinoless $0\nu2\beta$ decay with
different majorons emission were investigated with sensitivity
$T_{1/2}>10^{21}-10^{22}$ yr. New limits for the hypothetical
right-handed currents admixtures in weak interaction, the heavy
neutrino mass, and for the effective majoron-neutrino coupling
constants were set on the basis of the obtained $T_{1/2}$ limits.
Search for Lorentz-violating $2\nu2\beta$ decay of $^{116}$Cd was
realized for the first time resulting in the most stringent limit
on the Lorentz-violating parameter  $\mathring{a}
^{(3)}_{\mathrm{of}} \leq 4.0\times10^{-6}$ GeV.

\section{ACKNOWLEDGEMENTS}

The group from the Institute for Nuclear Research (Kyiv, Ukraine)
was supported in part by the program of the National Academy of
Sciences of Ukraine ``Fundamental research on high-energy physics
and nuclear physics (international cooperation)''. F.A.~Danevich
gratefully acknowledges support from the Jean d`Alembert
fellowship program (project CYGNUS) of the Paris-Saclay Excellence
Initiative, grant number ANR-10-IDEX-0003-02. A.S.~Barabash,
S.I.~Konovalov, V.N.~Shlegel and V.I.~Umatov were supported by
Russian Science Foundation (grant No. 18-12-00003).


\begin{thebibliography}{99}
 \bibitem{Tretyak:2002} V.I.~Tretyak, Yu.G.~Zdesenko, At. Data Nucl. Data Tables {\bf 80}, 83 (2002)
 \bibitem{Saakyan:2013}  R.~Saakyan, Annu. Rev. Nucl. Part. Sci. {\bf 63}, 503 (2013)
 \bibitem{Barabash0:2015} A.~S.~Barabash, Nucl. Phys. A {\bf935}, 52 (2015)
 \bibitem{Vergados:2016} J.D.~Vergados, H.~Ejiri, F.~\v{S}imkovic, Int. J. Mod. Phys. E {\bf25}, 1630007 (2016)
 \bibitem{Pas:2015} H.~P\"{a}s, W.~Rodejohann, New J. Phys. {\bf17}, 115010 (2015)
 \bibitem{Bilenky:2016} S.M.~Bilenky, C.~Giunti, Int. J. Mod. Phys. A {\bf30}, 1530001 (2015)
 \bibitem{Delloro:2016} S.~Dell'Oro, S.~Marcocci, M.~Viel, F.~Vissani, AHEP {\bf2016}, 2162659 (2016)
 \bibitem{Giuliani:2012} A.~Giuliani, A.~Poves, AHEP {\bf 2012}, 857016 (2012)
 \bibitem{Cremonesi:2014} O.~Cremonesi, M.~Pavan, AHEP {\bf2014}, 951432 (2014)
 \bibitem{Sarazin:2015} X.~Sarazin, J. Phys.: Conf. Ser. {\bf593}, 012006 (2015)
 \bibitem{Arnold:2015} R.~Arnold et al., Phys. Rev. D {\bf92}, 072011 (2015)
 \bibitem{Gando:2016} A.~Gando et al., Phys. Rev. Lett. {\bf117}, 082503 (2016)
 \bibitem{Albert:2018} J.B.~Albert et al., Phys. Rev. Lett. {\bf120}, 072701 (2018)
 \bibitem{Alduino:2018} C.~Alduino et al., Phys. Rev. Lett. {\bf 120}, 132501 (2018)
 \bibitem{Aalseth:2018} C.E.~Aalseth et al., Phys. Rev. Lett. {\bf120}, 132502 (2018)
 \bibitem{Agostini:2018} M.~Agostini et al., Phys. Rev. Lett. {\bf120}, 132503 (2018)
 \bibitem{Azzolini:2018} O.~Azzolini et al., Phys. Rev. Lett. {\bf 120}, 232502 (2018)
 \bibitem{Engel:2017} J.~Engel, J.~Menendez, Rep. Prog. Phys. {\bf80}, 046301 (2017)
 \bibitem{Barea:2015} J.~Barea, J.~Kotila, F.~Iachello, Phys. Rev. C {\bf91}, 034304 (2015)
 \bibitem{Kostensalo:2017} J.~Kostensalo, M.~Haaranen, and J.~Suhonen, Phys. Rev. C {\bf95}, 044313 (2017)
 \bibitem{Abad:1984} J.~Abad et al., J. Phys. Colloques {\bf45}, C3-147 (1984)
 \bibitem{Barabash:2007}  A.S.~Barabash et al., Nucl. Phys. B {\bf783}, 90 (2007)
 \bibitem{Diaz:2013} J.S.~Diaz, V.A.~Kostelecky, R.~Lehnert, Phys. Rev. D {\bf88}, 071902(R) (2013)
 \bibitem{Wang:2017} M.~Wang et al.,  Chin. Phys. C {\bf 41}, 030003 (2017)
 \bibitem{Simkovic:2013} F.~\v{S}imkovic, V.~Rodin, A.~Faessler, P.~Vogel, Phys. Rev. C {\bf87}, 045501 (2013)
 \bibitem{Vaquero:2013} N.L.~Vaquero, T.R.~Rodriguez, E.J.~Luis, Phys. Rev. Lett. {\bf111}, 142501 (2013)
 \bibitem{Hyvarinen:2015} J.~Hyv$\mathrm{\ddot{a}}$rinen, J.~Suhonen, Phys. Rev. C {\bf91}, 024613 (2015)
 \bibitem{Yao:2015} J.M.~Yao et al., Phys. Rev. C {\bf91}, 024316 (2015)
 \bibitem{Song:2017} L.S.~Song, J.M.~Yao, P.~Ring, J.~Meng, Phys. Rev. C {\bf95}, 024305 (2017)
 \bibitem{Meija:2016} J.~Meija et al., Pure Appl. Chem. {\bf 88}, 293 (2016)
 \bibitem{Danevich:2003a} F.A.~Danevich et al., Phys. Rev. C {\bf67}, 014310 (2003)
 \bibitem{Danevich:2003b} F.A.~Danevich et al., Phys. Rev. C {\bf68}, 035501 (2003)
 \bibitem{Belli:2007} P.~Belli et al., Phys. Rev. C {\bf76}, 064603 (2007)
 \bibitem{Belli:2008} P.~Belli et al., Eur. Phys. J. A {\bf36}, 167 (2008)
 \bibitem{Belli:2012} P.~Belli et al., Phys. Rev. C {\bf85}, 044610 (2012)
 \bibitem{Belli:2016} P.~Belli et al., Phys. Rev. C {\bf93}, 045502 (2016)
 \bibitem{CUPID} G.~Wang et al., arXiv:1504.03599v1 [physics.ins-det]
 \bibitem{Giuliani:2018} A.~Giuliani, F.A.~Danevich, V.I.~Tretyak, Eur. Phys. J. C {\bf78}, 272 (2018)
 \bibitem{Blachot:2010} J.~Blachot, Nucl. Data Sheets {\bf111}, 717 (2010)
 \bibitem{Ejiri:1995} H.~Ejiri et al., J. Phys. Soc. Japan {\bf64}, 339 (1995)
 \bibitem{Danevich:1995} F.A.~Danevich et al., Phys. Lett. B {\bf344}, 72 (1995)
 \bibitem{NEMO2:1995} R.~Arnold et al., JETP Lett. {\bf61}, 170 (1995)
 \bibitem{NEMO2:1996} R.~Arnold et al., Z. Phys. C {\bf72}, 239 (1996)
 \bibitem{Danevich:2000} F.A.~Danevich et al., Phys. Rev. C {\bf62}, 045501 (2000)
 \bibitem{NEMO3:2017} R.~Arnold et al., Phys. Rev. D {\bf95}, 012007 (2017)
 \bibitem{NEMO2*:2010} A.S.~Barabash, Phys. Rev. C {\bf81}, 035501 (2010)
 \bibitem{Barabash:1990} A.S.~Barabash, A.V.~Kopylov, V.I.~Cherehovsky, Phys. Lett. B {\bf249}, 186 (1990)
 \bibitem{Piepke:1994} A.~Piepke et al., Nucl. Phys. A {\bf577}, 493 (1994)
 \bibitem{Barabash:2013} A.S.~Barabash et al., Proc. 4-th Int. Conf. on Current Problems in Nucl. Phys. and At. Energy (NPAE-Kyiv2012), Kyiv, 2013, p. 353
 \bibitem{Poda:2014} D.V.~Poda et al., EPJ Web of Conferences {\bf65}, 01005 (2014)
 \bibitem{Polischuk:2015} O.G.~Polischuk et al., AIP Conf. Proc. {\bf1686}, 020017 (2015)
 \bibitem{Danevich:2016} F.A.~Danevich et al., J. Phys. Conf. Ser. {\bf718}, 062009 (2016)
 \bibitem{Polischuk:2017} O.G.~Polischuk et al., AIP Conf. Proc. {\bf1894}, 020018 (2017)
 \bibitem{Barabash:2011} A.S.~Barabash et al., JINST {\bf6}, P08011 (2011)
 \bibitem{Poda:2013} D.V.~Poda et al., Radiat. Meas. {\bf 56}, 66 (2013)
 \bibitem{Danevich:2013} F.A. Danevich et al.,  AIP Conf. Proc. {\bf1549}, 201 (2013)
 \bibitem{Gatti:1962} E.~Gatti, F.~De~Martini, Proceedings of the Conference on Nuclear Electronics. Vol. II, International Atomic Energy Agency, Vienna, 1962, p. 265.
 \bibitem{Fazzini:1998} T.~Fazzini et al., Nucl. Instrum. Meth. A {\bf410}, 213 (1998)
 \bibitem{Bardelli:2006} L.~Bardelli et al., Nucl. Instrum. Meth. A {\bf569}, 743 (2006)
 \bibitem{Barabash:2016} A.S.~Barabash et al., Nucl. Instrum. Meth. A {\bf833}, 77 (2016)
 \bibitem{Koskelo:1996} M.J.~Koskelo, W.C.~Burnett, P.H.~Cable, Radioact. Radiochem. {\bf7}, 18 (1996)
 \bibitem{Tretyak:2010} V.I.~Tretyak, Astropart. Phys. {\bf33}, 40 (2010)
 \bibitem{Barton:2000a} J.C.~Barton, J.A.~Edgington, Nucl. Instr. Meth. A {\bf443}, 277 (2000)
 \bibitem{Danevich:2001} F.A.~Danevich et al., Nucl. Phys. A {\bf694}, 375 (2001)
 \bibitem{EGS4} W.R.~Nelson, H.~Hirayama, D.W.O.~Rogers, The EGS4 Code System report SLAC-265, Stanford Linear Accelerator Center (1985)
 \bibitem{DECAY0} O.A.~Ponkratenko, V.I.~Tretyak, Yu.G.~Zdesenko, Phys. At. Nucl. {\bf63}, 1282 (2000); V.I. Tretyak, to be published.
 \bibitem{Doi:1985} M.~Doi, T.~Kotani, E.~Takasugi,  Prog. Theor. Phys. Suppl. {\bf83}, 1 (1985)
 \bibitem{Kotila:2012} J.~Kotila, F.~Iachello, Phys. Rev. C {\bf85}, 034316 (2012)
 \bibitem{Mirea:2015}  M.~Mirea, T.~Pahomi, S.~Stoica, Rom. Rep. Phys. {\bf67}, 872 (2015)
 \bibitem{Feldman:1998} G.J.~Feldman, R.D.~Cousins, Phys. Rev. D {\bf57}, 3873 (1998)
 \bibitem{Staudt:1990} A.~Staudt et al., Europhys. Lett. {\bf13}, 31 (1990)
 \bibitem{Pantis:1996} G.~Pantis et al., Phys. Rev. C {\bf53}, 695 (1996)
 \bibitem{R-parity} A.~Faessler et al., Phys. Rev. D {\bf58}, 115004 (1998)
 \bibitem{Faessler:2014} A.~Faessler et al., Phys. Rev. D {\bf90}, 096010 (2014)
 \bibitem{Barabash:2017} A.S.~Barabash, AIP Conf. Proc. {\bf1894}, 020002 (2017)
 \bibitem{Simkovic:2002} F.~\v{S}imkovic, A.~Faessler, Prog. Part. Nucl. Phys. {\bf48}, 201 (2002)
 \bibitem{Arnold:2000} R.~Arnold et al., Nucl. Phys. A {\bf678}, 341 (2000)
 \bibitem{Lattanzi:2013} M.~Lattanzi, S.~Riemer-S{\o}rensen, M.~T\'{o}rtola, J.W.F.~Valle, Phys. Rev. D {\bf88}, 063528 (2013)
 \bibitem{Kazanas:2004} D.~Kazanas, R.N.~Mohapatra, S.~Nasri, V.L.~Teplitz, Phys. Rev. D {\bf70}, 033015 (2004)
 \bibitem{Chikashige:1980} Y.~Chikashige, R.N.~Mohapatra, R.D.~Peccei, Phys. Rev. Lett. {\bf45}, 1926 (1980)
 \bibitem{Chikashige:1981} Y.~Chikashige, R.~Mohapatra, R.~Peccei, Phys. Lett. B {\bf98}, 265 (1981)
 \bibitem{Aulakh:1982} C.~Aulakh, R.~Mohapatra, Phys. Lett. B {\bf119}, 136 (1982)
 \bibitem{Gelmini:1981} G.~Gelmini, M.~Roncadelli, Phys. Lett. B {\bf99}, 411 (1981)
 \bibitem{Caso:1998} C.~Caso et al. (Particle Data Group), Eur. Phys. J. C {\bf3}, 1 (1998)
 \bibitem{Mohapatra:1991} R.N.~Mohapatra, P.B.~Pal, {\it Massive Neutrinos in Physics and Astrophysics} (World Scientific, Singapore, 1991)
 \bibitem{Berezhiani:1992} Z.G.~Berezhiani, A.Yu.~Smirnov, J.W.F.~Valle, Phys. Lett. B {\bf291}, 99 (1992)
 \bibitem{Mohapatra:1988} R.N.~Mohapatra, E.~Takasugi, Phys. Lett. B {\bf211}, 192 (1988)
 \bibitem{Burgess:1993} C.P.~Burgess, J.M.~Cline, Phys. Lett. B {\bf298}, 141 (1993)
 \bibitem{Burgess:1994} C.P.~Burgess, J.M.~Cline, Phys. Rev. D {\bf49}, 5925 (1994)
 \bibitem{Bamert:1995} P.~Bamert, C.P.~Burgess, R.N.~Mohapatra, Nucl. Phys. B {\bf449}, 25 (1995)
 \bibitem{Carone:1993} C.D.~Carone, Phys. Lett. B {\bf308}, 85 (1993)
 \bibitem{Mohapatra:2000} R.N.~Mohapatra, A.~Perez-Lorenzana, C.A.S.~Pires, Phys. Lett. B {\bf491}, 143 (2000)
 \bibitem{Kotila:2015} J.~Kotila, J.~Barea, F.~Iachello, Phys. Rev. C {\bf 91}, 064310 (2015)
 \bibitem{Kostelecky:2011} V.A.~Kostelecky, N. Russel, Rev. Mod. Phys. {\bf83}, 11 (2011)
 \bibitem{Tasson:2014} J.D.~Tasson, Rep. Prog. Phys. {\bf77}, 062901 (2014)
 \bibitem{Diaz:2014} J.S.~Diaz, Phys. Rev. D {\bf89}, 036002 (2014)
 \bibitem{Albert:2016} J.B.~Albert et al., Phys. Rev. D {\bf93}, 072001 (2016)
 \bibitem{Hirsch:1996} M.~Hirsch et al., Phys. Lett. B {\bf372}, 8 (1996)
 \bibitem{Beh69} H.~Behrens, J.~Janecke, {\it Numerical Tables for Beta-Decay and Electron Capture}, Berlin, Springer-Verlag, 1969.
 \bibitem{Pri59} H.~Primakoff, S.P.~Rosen, Rep. Prog. Phys. {\bf 22}, 121 (1959).

\end{thebibliography}
\end{document}